\documentclass[twocolumn]{openjournal}

\usepackage{ulem}
\usepackage{natbib}
\usepackage{amsmath}
\usepackage{orcidlink}
\usepackage{color}
\usepackage{hyperref}
\hypersetup{
  colorlinks=true,
  citecolor=blue,
}
\usepackage{threeparttable}
\usepackage{xspace}

\newcommand{\msun}{{\rm M}_\odot}

\newcommand{\kms}{{\rm km~s}^{-1}}
\newcommand{\dens}{{\rm M}_\odot~{\rm pc}^{-3}}
\newcommand{\fbg}{f_{\rm b,\ge 5\msun}}
\newcommand{\fbl}{f_{\rm b,< 5\msun}}
\newcommand{\tgw}{t_{\rm gw}}
\newcommand{\tgwacc}{t_{\rm gw,acc}}

\newcommand{\kshortp}{K_{\rm short,p}}
\newcommand{\mdark}{m_{\rm dark}}
\newcommand{\mvisi}{m_{\rm visible}}
\newcommand{\pout}{P}
\newcommand{\eout}{e}
\newcommand{\mbp}{m_{\rm p}}
\newcommand{\mpri}{m_1}
\newcommand{\msec}{m_2}
\newcommand{\pin}{P_{\rm in}}
\newcommand{\ein}{e_{\rm in}}
\newcommand{\Gaia}{Astrometric\xspace}
\newcommand{\gaia}{astrometric\xspace}

\begin{document}

\title{Compact Binary Formation in Open Star Clusters III:
  \\ Probability of Binary Black Holes Hidden in Gaia Black Hole
  Binary}

\author{Ataru Tanikawa\orcidlink{0000-0002-8461-5517}$^{1}$}
\author{Long Wang\orcidlink{0000-0001-8713-0366}$^{2,3}$}
\author{Michiko S. Fujii\orcidlink{0000-0002-6465-2978}$^{4}$}
\author{Alessandro A. Trani\orcidlink{0000-0001-5371-3432}$^{5}$}
\author{Toshinori Hayashi\orcidlink{0000-0003-0288-6901}$^{6}$}
\author{Yasushi Suto\orcidlink{0000-0002-4858-7598}$^{7}$}

\affiliation{$^1$Center for Information Science, Fukui Prefectural
  University, 4-1-1 Matsuoka Kenjojima, Eiheiji-cho, Fukui 910-1195,
  Japan}
\affiliation{$^2$School of Physics and Astronomy, Sun Yat-sen
  University, Daxue Road, Zhuhai, 519082, China}
\affiliation{$^3$CSST Science Center for the Guangdong-Hong Kong-Macau
  Greater Bay Area, Zhuhai, 519082, China}
\affiliation{$^4$Department of Astronomy, Graduate School of Science,
  The University of Tokyo, 7-3-1 Hongo, Bunkyo-ku, Tokyo 113-0033,
  Japan}
\affiliation{$^5$Niels Bohr International Academy, Niels Bohr
  Institute, Blegdamsvej 17, 2100 Copenhagen, Denmark}
\affiliation{$^6$Yukawa Institute for Theoretical Physics, Kyoto
  University, Kyoto 606-8502, Japan}
\affiliation{$^7$Research Institute, Kochi University of
  Technology, Tosa Yamada, Kochi 782-8502, Japan}
\email{Corresponding author: tanik@g.fpu.ac.jp}

\begin{abstract}

  The Gaia mission and its follow-up observations have discovered a
  few candidates of non-interacting single black holes (BHs) and
  visible stars, Gaia BH1, BH2, and BH3, collectively called ``\gaia
  BH binaries''. This paper investigates whether any of these
  candidates harbor binary BHs (BBHs), namely, whether any such
  candidates are previously undiscovered ``\gaia BBH
  triples''. Focusing on open star clusters, which are promising
  formation sites of \gaia BH binaries, we estimate the formation rate
  of \gaia BBH triples through gravitational $N$-body simulations. We
  find a competitively high formation efficiency of \gaia BBH triples
  ($\sim 10^{-6} {\rm M}_\odot^{-1}$ or $\sim 10$\% of \gaia BH
  binaries) in low-metallicity environments but no \gaia BBH triples
  in solar-metallicity environments. Most of the \gaia BBH triples in
  our simulations were dynamically stable for $10$ Gyrs, indicating
  that $\sim10$\% of \gaia BH binary candidates may indeed harbor
  inner BBHs if they originate from open star clusters in
  low-metallicity environments. \Gaia BBH triples can be distinguished
  from \gaia BH binaries through radial velocity follow-up of the
  tertiary star. According to the statistics of our simulated samples,
  a small percent of \gaia BH binary candidates should exhibit
  detectable radial-velocity modulations generated by inner BBHs. Such
  candidates preferentially exhibit ``outer'' orbital periods of
  $\gtrsim 10^3$ days and moderately high ``outer'' orbital
  eccentricities ($\gtrsim 0.7$). Our current result will strongly
  motivate the search for \gaia BBH triples in the upcoming Gaia Data
  Release 4 and Gaia Final Data Release.

\end{abstract}

\section{Introduction}
\label{sec:Introduction}

Stellar-mass black holes (BHs) with masses of $\sim 10$--$100 \msun$
are left behind after the evolution of massive stars. BHs are useful
physics laboratories for testing strong-gravity and high-energy
phenomena, and serve as astronomical probes for modeling the
evolutions of massive single and binary stars. Therefore astronomers
have searched for various BHs using complementary techniques:
gravitational microlensing for isolated BHs
\citep{2022ApJ...933L..23L, 2022ApJ...933...83S, 2025A&A...694A..94H},
X-rays \citep[see][for review]{2017hsn..book.1499C}, optical
spectroscopy \citep{2018MNRAS.475L..15G, 2019A&A...632A...3G,
  2022NatAs...6.1085S, 2022A&A...664A.159M}, and astrometry
\citep[see][for review]{2024NewAR..9801694E} for BH binaries
consisting of single BHs and single visible stars, and gravitational
waves (GWs) for binary BHs (BBHs) \citep{2023PhRvX..13d1039A}.

The Gaia mission \citep{2023A&A...674A...1G, 2023A&A...674A..34G},
which can potentially discover BH binaries with no significant mass
transfer, has thus far identified three BH binary candidates
(hereafter referred to as \gaia BH binaries): Gaia BH1
\citep{2023MNRAS.518.1057E, 2023AJ....166....6C}, Gaia BH2
\citep{2023ApJ...946...79T, 2023MNRAS.521.4323E}, and Gaia BH3
\citep{2024A&A...686L...2G}. The formation channel of \gaia BH
binaries remains an open and challenging question. Astrometric BH
binaries are characterized by long periods ($100$-$4000$ days) and
large eccentricities ($0.5$-$0.8$). Whereas Gaia BH3 may be formed in
an isolated binary evolution model \citep{2024OJAp....7E..38E,
  2024A&A...690A.144I}, Gaia BH1 and BH2 are incompatible with the
conventional binary evolution model\footnote{Neutron stars and white
dwarf binaries discovered from the Gaia database
\citep{2023ApJ...954....4G, 2023A&A...677A..11G, 2023SCPMA..6629512Z,
  2024MNRAS.527.11719, 2024OJAp....7E..27E, 2024OJAp....7E..58E} are
also incompatible with the conventional model.}.  Instead, Gaia BH1
and BH2 must be explained within the framework of isolated binary
models, requiring several modifications to the conventional model:
efficiency of common envelope ejection above the theoretical
expectation \citep{2022ApJ...931..107C, 2023MNRAS.518.1057E,
  2023MNRAS.521.4323E, 2023ApJ...953...52S}\footnote{However,
\cite{2022ApJ...937L..42H} claimed that common envelope ejection is
less efficient for \gaia BH binary progenitors.}, strong convective
tides on BH progenitors and moderate BH natal kicks
\citep{2024MNRAS.535.3577K}, high convective overshooting of the BH
progenitors \citep{2024MNRAS.535L..44G}, and strong wind that
suppresses the interaction between the BH progenitor and its companion
star \citep{2024A&A...692A.141K}.  Interestingly, we note that \gaia
BH binaries may be formed through triple star systems. For example,
\cite{2023MNRAS.518.1057E, 2023MNRAS.521.4323E} found that the merger
product of an inner massive binary maintains a nearly constant radius,
avoids common envelope evolution with a tertiary star, and finally
evolves to a BH with a visible companion such as a Gaia BH. Moreover,
\cite{2024ApJ...964...83G} reported that an inner binary evolves into
a Gaia BH-like system because the common envelope evolution is
eventually weakened by a 3-body secular or dynamical effect. \Gaia BH
binaries have also received substantial attention as indicators of BH
natal kicks \citep{2024PhRvL.132s1403V, 2025PASP..137c4203N,
  2025arXiv250416669W} and interactions in wide binaries with compact
objects \citep{2024ApJ...973...75C}.

The present paper explores whether BHs are dynamically formed in dense
star clusters. As members of the Galactic disk component
\citep{2023MNRAS.518.1057E, 2023MNRAS.521.4323E}, Gaia BH1 and Gaia
BH2 likely formed in open clusters \citep{2020PASJ...72...45S,
  2023MNRAS.526..740R, 2024MNRAS.527.4031T, 2024ApJ...965...22D}. Gaia
BH3 is located in the ED-2 stream \citep{2024A&A...687L...3B},
indicating its possible formation in a small globular cluster
\citep{2018ApJ...855L..15K, 2024A&A...688L...2M}. Alternatively, the
three BH binaries may have formed in massive star clusters
\citep{2025MNRAS.538..243F}.

\begin{figure}
  \centering
  \includegraphics[width=\columnwidth]{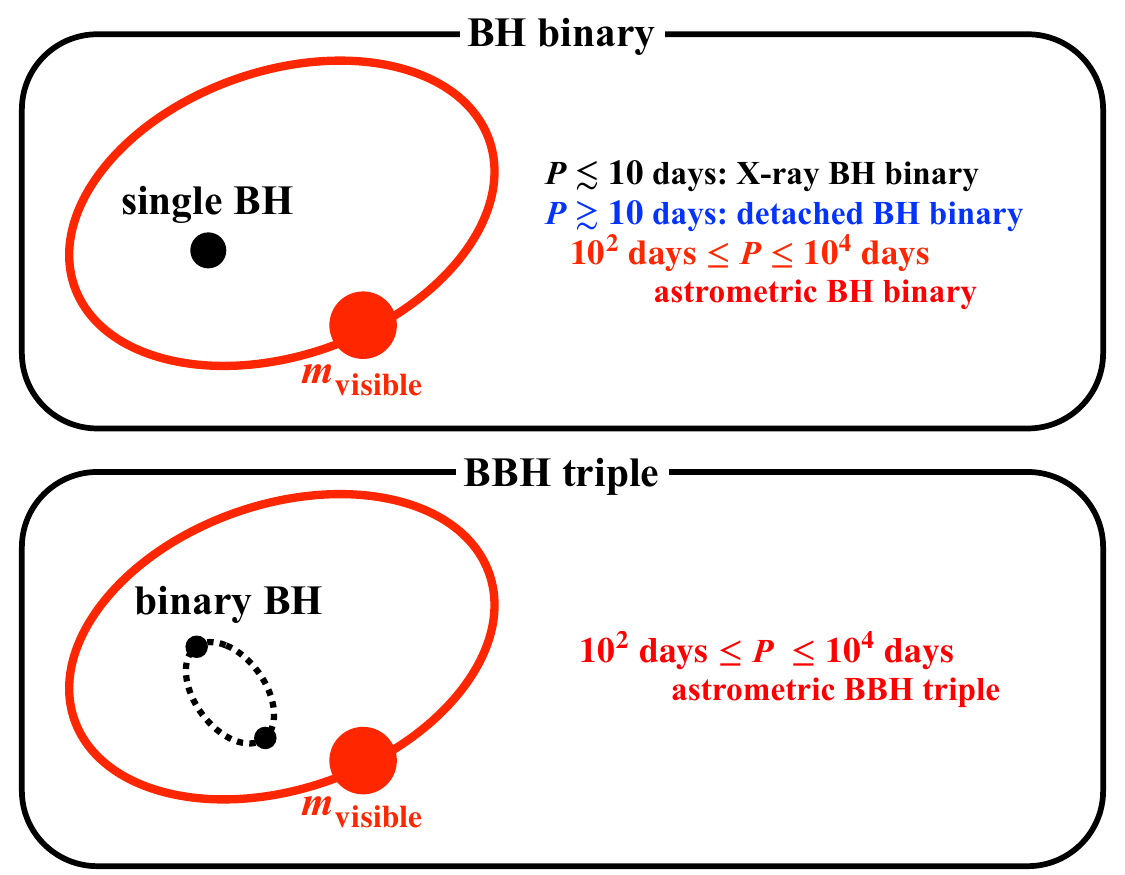}
  \caption{Descriptions of binary and triple star systems in the
    current analysis. A BH binary, consisting of one BH and one
    visible star, can be an X-ray or detached BH binary depending on
    its orbital period less than or more than approximately $10$ days,
    respectively \citep[see][and references
      therein]{2024OJAp....7E..58E}. Detached BH binaries contain
    \gaia BH binaries, defined as a subset of detached BH binaries
    with orbital periods of $10^2$-- $10^4$ days. The lower and upper
    limits depend on the observational cadence of Gaia ($60$ days) and
    the duration of Gaia's operation ($10$ years), respectively. A BBH
    triple consists of an inner BBH orbited by a tertiary visible
    star. An \gaia BBH triple defines a BBH triple with orbital period
    ranging from $10^2$ days to $10^4$ days, which will likely be
    initially identified as an \gaia BH binary (candidate) by the Gaia
    mission and confirmed as a triple after follow-up
    observations. The stellar orbit is simply referred to as the
    ``orbit'' because it corresponds to the orbit of the BH binary
    candidate.}
  \label{fig:terminology}
\end{figure}

One intriguing and fascinating question is whether any \gaia BH binary
is actually a triple system in which the central dark object is a BBH
(not a single BH) orbited by a visible tertiary star
\citep{2022ApJ...939...81H, 2023ApJ...958...26H}. Specifically, we
define a possible population of BBH triples with outer-orbital periods
ranging from $10^2$ to $10^4$ days as ``\gaia BBH triples''. After
identifying these triples as \gaia BH binaries, we detect the binarity
of the central dark object (initially assumed as a single BH) through
radial velocity follow-ups of the tertiary star. Our terminology is
summarized in Figure \ref{fig:terminology}, and section
\ref{sec:terminology}.

\cite{2024PASP..136a4202N} imposed strong constraints on the binarity
of the central object in Gaia BH1, whereas Gaia BH3 contains a massive
BH ($\sim 33\;\msun$), \citep{2024A&A...686L...2G}, which may imply a
BBH. However, as no observational constraints have been placed on Gaia
BH3, no BBH triple has been discovered thus far. Nevertheless, we
emphasize that any existing \gaia BBH triples are probably detectable
at the current precision of radial-velocity measurements
\citep{2020ApJ...890..112H, 2020ApJ...897...29H,
  2022PhRvD.106l3010L}. To this end, we numerically estimate the
fraction of \gaia BBH triples formed in open clusters.

Theoretical insights can be gained by investigating the dynamical
formation channel of \gaia BBH triples in open clusters, which is
simply dictated by gravity between the BBH progenitors and tertiary
stars. In reality, \gaia BBH triples can also be formed through
isolated triple evolution \citep{2009ApJ...697.1048P,
  2012ApJ...760...99P, 2014ApJ...794..122M, 2016ARA&A..54..441N,
  2017ApJ...841...77A, 2019MNRAS.488.2480R, 2020A&A...640A..16T,
  2022A&A...661A..61T, 2022ApJ...925..178H, 2022MNRAS.516.1406S,
  2023A&A...678A..60K, 2023ApJ...955L..14S, 2023MNRAS.518.1057E,
  2023MNRAS.521.4323E, 2024ApJ...964...83G}. Considering the
inevitable uncertainties in the hydrodynamical interactions between
the BBH progenitor and tertiary star
\citep[e.g.][]{2021MNRAS.502.4479H}, we interpret the formation rate
in open clusters as a lower limit on the \gaia BBH triple formation
rate.

The remainder of this paper is structured as follows. Section
\ref{sec:Method} presents our numerical method for modeling the
evolution of \gaia BH binaries and \gaia BBH triples in open
clusters. Section \ref{sec:Results} displays our main results and
Section \ref{sec:Distinguishability} discusses the detectability of
\gaia BBH triples in our simulation. Section
\ref{sec:DiscussionAndConclusion} concludes the paper and discusses
future implications of our work.

\section{Method}
\label{sec:Method}

After summarizing our terminology (subsection \ref{sec:terminology}),
we present our method for simulating open clusters (subsection
\ref{sec:petar}) and our approach for following the secular evolution
of \gaia BBH triples (subsection \ref{sec:okinami}).

\subsection{Terminology}
\label{sec:terminology}

Our terminology is summarized in Figure \ref{fig:terminology}. A
binary system consisting of one BH and one visible star is called a
``BH binary''. A visible star orbiting a BBH forms a triple system
called a ``BBH triple'' characterized by two independent
attributes. First, a BBH can reside inside the triple (an ``inner
BBH'') or outside the triple (a ``non-inner BBH'' or simply a
``BBH''). Second, a BBH can merge within 10 gigayears (Gyrs) (a
``merging BBH'') or not merge within 10 Gyrs (a non-merging BBH). The
timescale of a BBH merger is calculated as the GW radiation timescale
($\tgw$) using Peters' formula \citep{1963PhRv..131..435P,
  1964PhRv..136.1224P}. Because $\tgw$ is calculated using the orbital
parameters of a BBH at the end of the simulation, the BBH is defined
as merging if $\tgw < 10$ Gyrs, not if $\tgw < 13.8$ Gyrs (the Hubble
time). At the end of our simulation, at least 1 Gyr has passed since
the birth of the universe and $10$ Gyrs is a well-rounded number.

When we mention the orbit, orbital period, and orbital eccentricity of
a BBH triple, they mean outer binary orbits, periods, and
eccentricities, respectively, if unstated. This is because they
correspond to orbit, orbital period, and orbital eccentricity of a BH
binary. The inner and outer orbits of BBH triples are referred to as
``inner orbit'' and ``outer orbit'', respectively. The BHs in BH
binaries and inner BBHs in BBH triples are collectively called ``dark
objects''. The parameters of the BH binaries and BBH triples are
summarized in Table \ref{tab:notation}. Here, $\mdark$, $\mvisi$,
$\pout$, and $\eout$ denote the dark-object masses, visible star
masses, orbital periods, and orbital eccentricities, respectively, of
the \gaia BH binaries and \gaia BBH triples, whereas $\mpri$, $\msec$,
$\pin$, and $\ein$ express the primary BH masses, secondary BH masses,
orbital periods, and orbital eccentricities of the inner BBHs of the
\gaia BBH triples, respectively.

\begin{table*}
  \centering
  \begin{threeparttable}
  \caption{Parameters of the BH binaries and BBH
    triples in the present study} \label{tab:notation}
  \begin{tabular}{lll}
    \hline
    Parameter & BH binary & BBH triple\\
    \hline
    $\mdark$ & BH mass & Inner BBH mass\\
    $\mvisi$ & Visible star mass & Tertiary visible star mass\\
    $\pout$  & Orbital period & Outer-orbital period\\
    $\eout$  & Orbital eccentricity & Outer-orbital eccentricity\\
    \hline
    $\mpri$  & N/A & Primary BH mass of inner BBH\\
    $\msec$  & N/A & Secondary BH mass of inner BBH\\
    $\pin$   & N/A & Inner orbital period\\
    $\ein$   & N/A & Inner orbital eccentricity\\
    \hline
  \end{tabular}
  \end{threeparttable}
\end{table*}

\begin{table*}
  \centering
  \begin{threeparttable}
  \caption{Classification of related
    objects} \label{tab:classification}
  \begin{tabular}{lccc}
    \hline
    Name & Orbital period ($P$) & Visible star type & Visible star mass ($\mvisi$) \\
         & [day] & & [$\msun$]                    \\
    \hline
    BH binary & Any & Visible star       & Any        \\
    \Gaia BH binary        & $10^2$--$10^4$ & Main-sequence star & $\leq 1.1$ \\
    Quasi-\gaia BH binary  & $\leq 10^6$      & Main-sequence star & Any        \\
    BBH triple & Any & Visible star       & Any        \\
    \Gaia BBH triple       & $10^2$--$10^4$ & Main-sequence star & $\leq 1.1$ \\
    Quasi-\gaia BBH triple & $\leq 10^6$      & Main-sequence star & Any        \\
    \hline
  \end{tabular}
  \begin{tablenotes}
  \item ``Visible stars'' incorporate main-sequence stars, post main-sequence
    stars, and helium stars. They exclude white dwarfs, neutron stars, and BHs.
  \item Quasi-\gaia BH binaries (quasi-\gaia BBH triples) exclude
    \gaia BH binaries (\gaia BBH triples).
  \end{tablenotes}
  \end{threeparttable}
\end{table*}

The $\tgw$ of the inner BBH of an \gaia BBH triple is then expressed
as
\begin{align}
  \tgw = \frac{5}{256} \frac{c^5}{G^3} \frac{a_{\rm
      in}^4}{m_1m_2(m_1+m_2)}
  \frac{(1-\ein^2)^{7/2}}{(1+\frac{73}{24}\ein^2+\frac{37}{96}\ein^4)},
\end{align}
where $a_{\rm in }$ is the semi-major axis of the inner BBH, and $G$
and $c$ are the gravitational constant and speed of light,
respectively. \cite{2023MNRAS.524..426I} pointed out that this
definition may underestimate the GW radiation timescale at high
$\ein$.  To detect changes in our results, we also define $\tgwacc$ as
follows:
\begin{align}
  \tgwacc &= \frac{5}{256} \frac{c^5}{G^3} \frac{a_{\rm
      in}^4}{m_1m_2(m_1+m_2)} (1-\ein^2)^{7/2} \nonumber \\
  & \times (1+0.27\ein^{10}+0.33\ein^{20}+0.2\ein^{100})
\end{align}
\citep{2021RNAAS...5..223M}. According to \cite{2023MNRAS.524..426I},
$\tgwacc$ can estimate the GW radiation timescale to within $3$\% of
the GW radiation timescale calculated (more accurately) by numerical
integration.

Table \ref{tab:classification} classifies the BH binaries and BBH
triples by their orbital periods and visible star masses. The main
targets of this paper, namely, the \gaia BH binaries and \gaia BBH
triples, satisfy the following conditions:
\begin{enumerate}
\item $10^2 \le \pout/{\rm day} \le 10^4$. The lower and upper limits
  of $P$ depend on the observational cadence of Gaia ($60$ days), and
  the operation duration of Gaia ($10$ years), respectively. Orbital
  solutions can be found for binaries with orbital periods of
  approximately double the operation duration
  \citep[e.g.][]{2014A&A...563A.126L, 2019AJ....158....4O}.
\item Visible stars are main-sequence stars. Our definition of \gaia
  BH binaries supposedly encompasses Gaia BH2 and BH3, which both
  contain stars outside the main sequence (namely, red giants)
  along with visible stars that have resided in the main-sequence
  stars over a long period. We interpret that these visible stars in
  Gaia BH2 and BH3 have evolved off the main sequence by chance.
\item $\mvisi \le 1.1 \msun$. Considering the visible star masses of
  Gaia BH1 ($0.93 \msun$), Gaia BH2 ($1.07 \msun$), and Gaia BH3
  ($0.76 \msun$), we set a much smaller $\mvisi$ than the BH
  progenitor mass ($\gtrsim 30\;\msun$). Then, \gaia BH binary
  formation may be dominated by the dynamical interactions in dense
  star clusters, not by isolated binary evolution. A similar outcome
  can be obtained by setting $\mvisi \ll 30\;\msun$. For example, the
  formation efficiencies of \gaia BH binaries and BBH triples increase
  by only a few 10 \%, if we change the upper limit of $\mvisi$ to
  $2.5\;\msun$, where $2.5\;\msun$ stars do not yet evolve to white
  dwarfs within 1 Gyrs\footnote{An increase of several 10 \% should be
  insignificant because the formation efficiency is influenced by many
  uncertainties that either increase or decrease the formation
  efficiencies; for example, the initial cluster conditions such as
  the initial binary fractions and mass
  densities. \label{note:few10percent}}. This is because $>
  1.1\;\msun$ stars are less numerous than $\le 1.1\;\msun$ stars in
  the top-light initial mass function (IMF) adopted in our study.
\item All \gaia BH binaries and \gaia BBH triples are located outside
  of clusters. In fact, our cluster models are tidally disrupted at 1
  Gyr, meaning that no \gaia BH binaries or \gaia BBH triples can
  reside inside clusters.
\end{enumerate}

We also define systems with $\pout \leq 10^6$ days as ``quasi-\gaia BH
binaries'' or ``quasi-\gaia BBH triples''. Their BH companion and
tertiary star are main-sequence stars. They do not include \gaia BH
binaries nor \gaia BBH triples.

\subsection{$N$-body simulations of open clusters}
\label{sec:petar}

The open clusters were dynamically evolved through gravitational
$N$-body simulations as described in \cite{2024OJAp....7E..39T}. Here,
we briefly describe the methods relevant to our study. Simulations
were performed in the $N$-body code PETAR \citep{2020MNRAS.497..536W},
which is highly optimized by FDPS \citep{2016PASJ...68...54I,
  2020PASJ...72...13I}. Binary and close-encounter orbits are solved
with high accuracy using the SDAR code
\citep{2020MNRAS.493.3398W}. The mass model of the Galaxy was the {\tt
  MWPotential2014} model constructed in the GALPY code
\citep{2015ApJS..216...29B}. The PETAR code is equipped with the BSE
code \citep{2000MNRAS.315..543H, 2002MNRAS.329..897H,
  2020A&A...639A..41B}, which handles single and binary star
evolutions. As the supernova model, we adopted the rapid model
\citep{2012ApJ...749...91F} with the correction of moderate pair
instability and pulsational pair-instability supernovae
\citep{2020A&A...636A.104B}. The velocities of the natal kicks,
delivered to neutron stars (NSs) and BHs at birth, followed a single
Maxwellian distribution with $265$ $\kms$ \citep{2005MNRAS.360..974H}
and were decreased by $1-f_{\rm fb}$ where $f_{\rm fb}$ is the
fraction of fallback mass \citep{2012ApJ...749...91F}. The fallback
effect largely reduced the velocity of BH natal kicks, consistent with
the observational constraints in \cite{2023MNRAS.525.1498Z} and
\cite{2025PASP..137c4203N}. The BSE code handles common envelope
evolution with the $\alpha$ formalism
\citep{1984ApJ...277..355W}. Common envelope evolution begins when the
ratio of the donor-star mass to the accretor-star mass is $>3$ for a
donor star with a radiative envelope and $>0.362+1/[3(1-m_{\rm
    d,c}/m_{\rm d})]$ for a donor star with a convective
envelope. Here, $m_{\rm d}$ and $m_{\rm d,c}$ denote the mass and core
mass of the donor-star, respectively. In our simulations, we chose
$\alpha_{\rm CE}=1$ and adopted the $\lambda_{\rm CE}$ in
\cite{2014A&A...563A..83C}, which largely reduce the binary formation
efficiency of \gaia BHs from those of primordial binaries. This is
because BH binaries tend to have orbital periods of $<100$ days if
they experience common envelope evolution.

\begin{table}
  \centering
  \begin{threeparttable}
  \caption{Summary of cluster models.} \label{tab:ic}
  \begin{tabular}{lcccc}
    \hline
    Name & $Z$ & $M$       & $\rho$    & $\fbl$ \\
         & & [$\msun$] & [$\dens$] &        \\
    \hline
    $Z=0.02$ & $0.02$ & $1000$ & $20$ & $20$~\% \\
    $Z=0.01$ & $0.01$ & $1000$ & $20$ & $20$~\% \\
    $Z=0.005$ & $0.005$ & $1000$ & $20$ & $20$~\% \\
    $Z=0.002$ & $0.002$ & $1000$ & $20$ & $20$~\% \\
    $Z=0.0002$ & $0.0002$ & $1000$ & $20$ & $20$~\% \\
    \hline
  \end{tabular}
  \begin{tablenotes}
  \item Note: The 4th column $\rho$ records the initial density within
    the initial half-mass radius. The 5th column $\fbl$ is the initial
    binary fraction of primordial binaries with primary masses $<
    5\msun$.
  \end{tablenotes}
  \end{threeparttable}
\end{table}

The initial conditions of our cluster models are summarized in Table
\ref{tab:ic}. At the initial time, the mass ($M$) and mass density
within the half-mass radius ($\rho$) of each cluster were $10^3
\msun$, and $20 \dens$, respectively, consistent with the observed
Milky Way open clusters \citep{2010ARA&A..48..431P,
  2019ARA&A..57..227K}. Note that those clusters have already become
diffuse through star-evolution mass loss and two-body
relaxation. Their initial distributions in phase space follow the
Plummer model \citep{1911MNRAS..71..460P} without primordial mass
segregation or fractality. Each cluster circularly orbits at a
distance of $8$ kpc from the Galactic center. Depending on the primary
star masses ($\mbp$), the initial binary fractions of $\mbp \ge 5
\msun$ ($\fbg$) and $\mbp < 5 \msun$ ($\fbl$) were $100$ and $20$ \%,
respectively. Kroupa's IMF was adopted for the single stars and
primary stars in binaries \citep{2001MNRAS.322..231K}. The initial
conditions of the binary parameters, such as the secondary star
masses, binary periods, and eccentricities, were those in
\cite{2012Sci...337..444S} for $\mbp \ge 5 \msun$ and those in
\cite{1995MNRAS.277.1491K, 1995MNRAS.277.1507K} with modifications of
\cite{2017MNRAS.464.4077B} for $\mbp < 5 \msun$\footnote{Note that we
adopted $\fbl=20$\%, whereas \cite{1995MNRAS.277.1491K,
  1995MNRAS.277.1507K} adopted $\fbl=100$\%. If $\fbl > 20$\% in
reality, we underestimate the formation efficiency of \gaia BH
binaries because such binaries form when a single BH dynamically
interacts with a primordial binary and captures one of its
members. Therefore, the formation efficiency is proportional to $\fbl$
\citep{2024OJAp....7E..39T} and the formation efficiency of \gaia BBH
triples may show a similar proportionality relation. Nevertheless, we
reduce the formation efficiencies of \gaia BH binaries and \gaia BBH
triples by a small factor (5) because $\fbl$ can reach $100$\% in
principle. Furthermore, supposing that the formation efficiencies of
\gaia BH binaries and \gaia BBH triples are proportional to $\fbl$, we
expect that changing $\fbl$ will not change the probability of \gaia
BBH triples to \gaia BH binaries.}. We note that the mass ratio of
primordial binaries with $\mbp \ge 5 \msun$ ranges from $0.1$ to
$1$\footnote{This choice substantially influences the formation of
\gaia BH binaries from primordial binaries
\citep{2023MNRAS.526..740R}. Specifically, if we decrease the lower
limit of the mass ratio of the primordial binaries, we largely
increase the \gaia BH binary formation efficiency.}. We varied the
metallicity as $Z=0.02, 0.01, 0.005, 0.002$, and $0.0002$ and
generated $10^4$ cluster models for each metallicity, obtaining
clusters with a total mass of $5 \times 10^7 \msun$.  All cluster
models were constructed in MCLUSTER \citep{2011MNRAS.417.2300K}. The
cluster simulations were terminated at $1$ Gyr. At that time, all
clusters had become completely disrupted.

\cite{2024OJAp....7E..39T} have treated other cluster models with
cluster masses ($M=200$, $500$, and $2000$ $\msun$), cluster mass
densities ($\rho=2$ and $200 \dens$), and initial binary fractions for
$\mbp < 5 \msun$ ($\fbl=0$ and $50$ \%). As these cluster models with
high metallicity ($Z=0.02$) cannot form \gaia BBH triples, their
exclusion from the present paper does not affect our
conclusions. Therefore, we consider that the $Z=0.02$ cluster model
with $M=10^3 \msun$, $\rho=20 \dens$, and $\fbl=20$ \% is sufficiently
representative of high-metallicity clusters.

The formation efficiencies of \gaia BBH triples in low-metallicity
environments likely depend on the cluster parameters, as observed for
\gaia BH binaries in solar-metallicity environments
\citep{2024OJAp....7E..39T}. This proposition is reasonable because
\gaia BBH triples are formed similarly to \gaia BH binaries; that is,
a BH (or BBH) closely interacts with a main-sequence binary and
captures one of its members. However, a thorough survey of cluster
models with various metallicities, cluster masses, cluster densities,
initial binary fractions, and other parameters is prohibitively
time-intensive and beyond the scope of this paper.

Considering different galactocentric distances and eccentric orbits of
the open clusters, the formation efficiencies of \gaia BBH triples can
also depend on the galactic tidal field. However, we consider that the
galactic tidal field in our cluster models is reasonable because the
currently discovered \gaia BH binaries are $\lesssim 1$ kpc from the
Sun, implying that their galactocentric distances ($7$--$9$ kpc) are
similar to those in our cluster models. Therefore, our results can be
directly compared with observations of \gaia BH binaries. Moreover, if
their time-averaged galactocentric distances are fixed, clusters on
eccentric orbits evolve similarly to those on circular orbits
\citep{2003MNRAS.340..227B}. We thus suppose that clusters forming
\gaia BH binaries and \gaia BBH triples at a galactocentric distance
of $\sim 8$ kpc evolve similarly, regardless of whether their orbits
are circular or eccentric.

From our clusters aged $1$ Gyr, we extracted the \gaia BH binaries and
\gaia BBH triples. The conditions of the \gaia BH binaries and \gaia
BBH triples are detailed in subsection \ref{sec:terminology}.

\subsection{Three-body simulations of \gaia BBH triples}
\label{sec:okinami}

To assess the stability of the \gaia BBH triples over $10$ Gyrs, we
followed the secular evolutions of the \gaia BBH triples in the
\textsc{okinami} code \citep{2023IAUS..362..404T}, which numerically
solves a stable hierarchical triple system by integrating the
equations of motion in Delunay elements derived from a three-body
Hamiltonian expanded at the octupole-level interaction and averaged
over the mean anomalies of the inner and outer binaries
\citep{naoz2016ARA&A..54..441N,2023MNRAS.522..937T}.  The equations
are integrated using an 8th-order Dormand--Prince scheme with an
adaptive step size. Post-Newtonian corrections of order 1, 2, and 2.5
were applied using the midpoint step method described in
\cite{2008AJ....135.2398M}.

\section{Results}
\label{sec:Results}

This section presents our simulation results. Subsection
\ref{sec:FormationEfficiency} compares the formation efficiencies of
the \gaia BBH triples and \gaia BH binaries. Subsection
\ref{sec:GaiaBBHtriple} describes the orbital properties of the \gaia
BBH triples, which can assist future searches for \gaia BBH
triples. Subsection \ref{sec:InnerBBH} comprehensively investigates
the inner BBHs in the \gaia BBH triples and compares them with
non-inner and inner BBHs in quasi-\gaia BBH triples.

\subsection{Formation efficiency of \gaia BBH triples}
\label{sec:FormationEfficiency}

Among $48$ \gaia BBH triples found in our cluster models aged $1$ Gyr,
$0$, $1$, $6$, $20$, and $21$ triples were assigned to the $Z=0.02$,
$0.01$, $0.005$, $0.002$, and $0.0002$ models, respectively. All inner
BBHs were formed from primordial binaries and captured visible stars
through interactions with main-sequence binaries. Quadruple systems
consisting of a BBH and another binary with at least one visible star
were not observed.

\begin{figure}
  \includegraphics[width=\columnwidth]{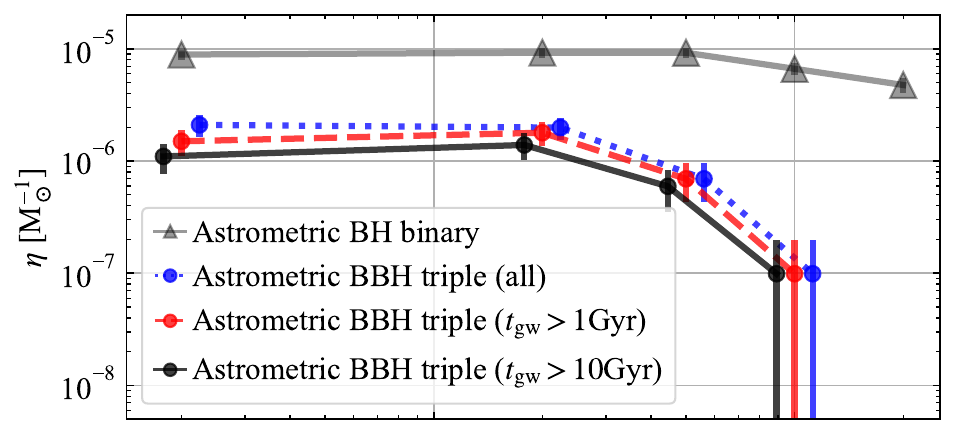}
  \includegraphics[width=\columnwidth]{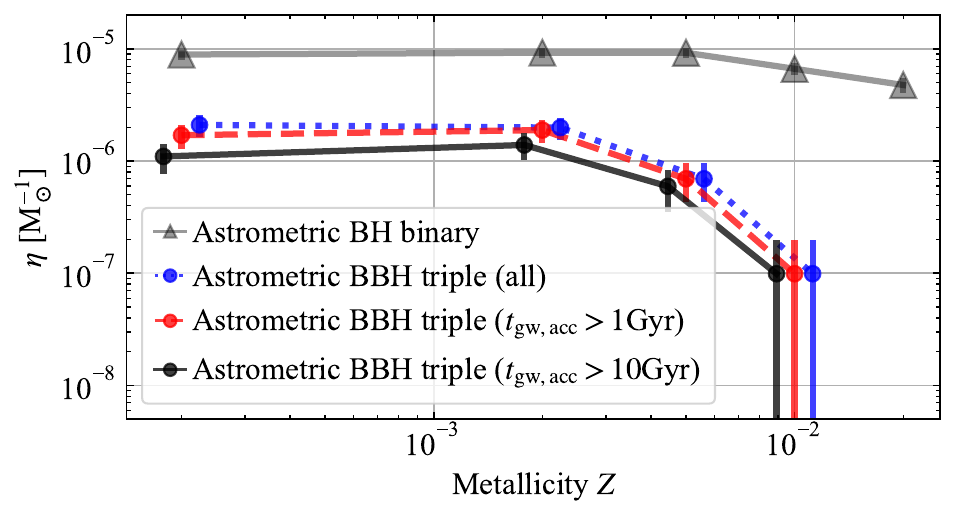}
  \caption{Formation efficiencies of \gaia BBH triples at $1$ Gyr as
    functions of cluster metallicity. The upper panel plots the
    formation efficiencies of all \gaia BBH triples (circles and
    dotted curves) and the \gaia BBH triples with $\tgw>1$ and
    $\tgw>10$ Gyrs (dashed and solid curves, respectively) In the
    bottom panel, $\tgw$ is replaced with $\tgwacc$. For reference,
    the formation efficiencies of \gaia BH binaries are also plotted
    (triangles and gray solid curve). To enhance visibility, the
    circles along the dotted and solid curves are shifted by $0.05$
    dex rightward and leftward, respectively. The error bars are the
    square roots of the number of corresponding objects, assuming that
    the formations of \gaia BH binaries and \gaia BBH triples follow a
    Poisson process. No points are plotted at $Z=0.02$ because no
    \gaia BBH triples were formed at this metallicity.}
  \label{fig:etaGaiabbh}
\end{figure}

The upper panel of Figure \ref{fig:etaGaiabbh} plots the formation
efficiencies of \gaia BH binaries and \gaia BBH triples as functions
of cluster metallicity. The formation efficiencies of \gaia BH
binaries are $\sim 10^{-6}$--$10^{-5} \msun^{-1}$, as also shown in
\cite{2024OJAp....7E..39T}. The formation efficiencies of all \gaia
BBH triples strongly depend on cluster metallicity, being $\sim
10^{-6} \msun^{-1}$ in low-metallicity environments ($Z \le 0.005$),
$10^{-7} \msun$ at $Z=0.01$, and $0$ at $Z=0.02$. This metallicity
dependence results from the formation efficiencies of short-period
BBHs in isolated binaries; short-period BBHs are commonly formed in
low-metallicity environments but sparsely formed in solar-metallicity
environments as shown later \citep[see
  also][]{2020ApJ...898..152S}. The inner BBHs in these \gaia BBH
triples are formed from primordial binaries (or effectively isolated
binaries in open clusters) and will be more intensively examined in
subsection \ref{sec:InnerBBH}.

\begin{figure}
  \includegraphics[width=\columnwidth]{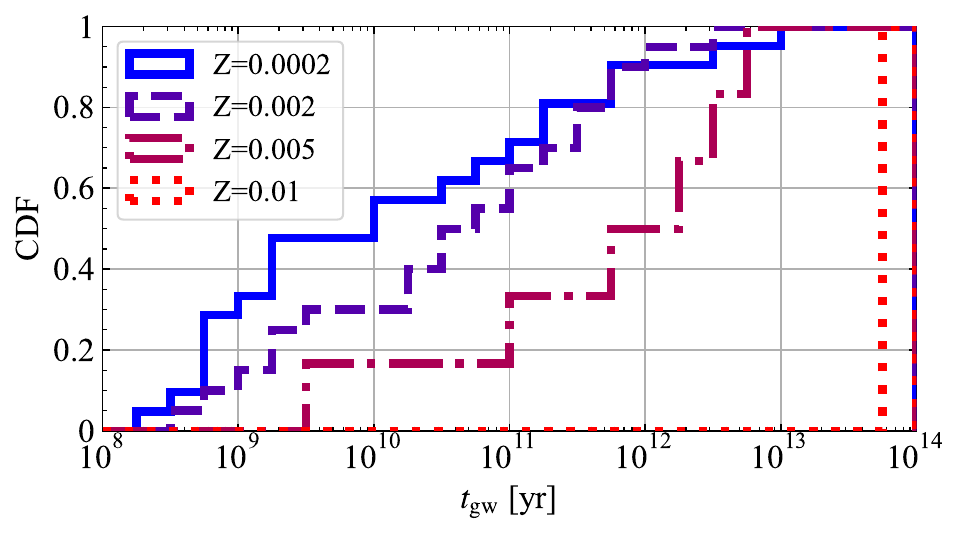}
  \caption{Cumulative distributions of GW radiation timescale ($\tgw$)
    of inner BBHs of \gaia BBH triples in the $Z=0.0002$, $0.002$,
    $0.005$ and $0.01$ models (no \gaia BBH Triples were found in the
    $Z=0.02$ model).}
  \label{fig:mergertimeGaiabbh}
\end{figure}

Long-term survival of all \gaia BBH triples is not expected. It is
probably difficult to discover some of their inner BBHs with $\tgw <
1$ Gyrs because all currently discovered \gaia BH binaries are aged
beyond $1$ Gyr \citep{2023MNRAS.518.1057E, 2023MNRAS.521.4323E,
  2024A&A...686L...2G}. In contrast, inner BBHs with $\tgw > 1$ Gyr
constitute a large fraction of \gaia BBH triples.  Figure
\ref{fig:mergertimeGaiabbh} shows the cumulative $\tgw$ distributions
of the inner BBHs in the different metallicity models. In the
$Z=0.01$, $Z=0.005$, $Z=0.002$, and $Z=0.0002$ models, the proportions
of \gaia BBH triples with $\tgw > 1$ Gyr were $100$\%, $100$\%,
$90$\%, and $70$\% , respectively, and the proportions of non-merging
BBHs categorized among the inner BBHs are $100$\%, $80$\%, $70$\%, and
$50$\%, respectively. Assuming that only \gaia BBH triples with
non-merging BBHs can be discovered, the formation efficiencies of such
triples is $\sim 10^{-6} \msun^{-1}$ in the $Z \le 0.005$ environments
and $\sim 10^7 \msun^{-1}$ in the $Z=0.01$ environment. In other
words, we can expect that one \gaia BBH triple is included for every
10 \gaia BH binaries in the $Z \le 0.005$ environments like Gaia BH3,
since the formation efficiency of \gaia BH binaries in the $Z \le
0.005$ environments is $\sim 10^{-5} \msun^{-1}$.

The similar formation efficiencies of \gaia BBH triples with $\tgw$
and $\tgwacc$ (see the lower panel of Figure \ref{fig:etaGaiabbh}) is
explained by the low $\ein$ of the inner BBHs of \gaia BBH triples. As
described in \cite{2023MNRAS.524..426I}, $\tgw$ and $\tgwacc$ are
comparable when $\ein$ is small. Here, $\ein$ is low because the inner
BBHs were effectively formed through isolated binary evolution.

More details are displayed in Figure \ref{fig:mergertimeGaiabbh}. The
cumulative distributions exhibit two distinct features. First, no
\gaia BBH triples with $\tgw < 10^8$ yrs are seen because their
survival chances diminish at the end of our simulations, when the
cluster age is $1$ Gyrs. Second, because the cumulative distributions
increase proportionally to the logarithm of $\tgw$, the $\tgw$
distribution is logarithmically flat, similarly to that of double
compact binaries \citep[e.g.][]{2008PASJ...60.1327T}. It appears that
the $\tgw$ timescales of the inner BBHs in \gaia BBH triples are
reduced in low-metallicity environments. A Kolmogorov--Smirnov test
revealed no significant differences among the $\tgw$ distributions of
the $Z=0.0002$ and other models; the P-values are $>0.07$.

The formation efficiencies of \gaia BBH triples with non-merging inner
BBHs are $\sim 10$\%, $\sim 1$\%, and $0$\% of the \gaia BH binaries
in the $Z \le 0.005$, $Z=0.01$, and $Z=0.02$ environments,
respectively, and may represent upper limits. We predict the lifetimes
of \gaia BBH triples based only on their $\tgw$ values, but the actual
lifetimes can be shortened by the Kozai--Lidov mechanism
\citep{1924MNRAS..84..665V, 1962AJ.....67..591K, 1962P&SS....9..719L},
which may accelerate the merger process of their inner BBHs
\citep{2013PhRvL.111f1106S, 2016ApJ...831..187A, 2017ApJ...841...77A,
  2017ApJ...836...39S, 2018ApJ...863....7R, 2018A&A...610A..22T,
  2020ApJ...903...67M, 2021ApJ...907L..19V, 2021A&A...650A.189A,
  2022ApJ...937...78M, 2022MNRAS.511.1362T, 2024MNRAS.533.2262L}. To
assess whether the Kozai--Lidov mechanism can shorten the merger time
of the inner BBHs in \gaia BBH triples to within 10 Gyrs, we ran
simulations in the OKINAMI code. As no mergers within $10$ Gyrs
appeared in the inner BBHs of \gaia BBH triples with $\tgw$ exceeding
$10$ Gyrs, we concluded that the formation efficiencies of \gaia BBH
triples are those presented in Figure \ref{fig:etaGaiabbh}, and that
$10$\% of the \gaia BH binaries in $Z \le 0.005$ environments can be
\gaia BBH triples. Note that Gaia BH3 should be formed in $Z \le
0.005$ environments.

\begin{figure}
  \includegraphics[width=\columnwidth]{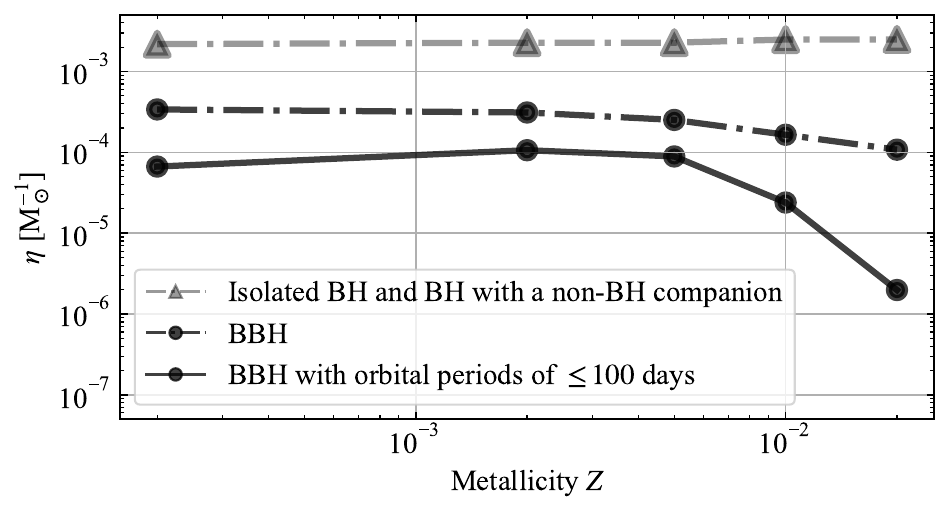}
  \caption{Formation efficiencies of all BBHs, BBHs with orbital
    periods of $\leq 10^2$ days, and sum of the isolated BHs and BHs
    with non-BH companions.}
  \label{fig:countBbh}
\end{figure}

As described above, the formation efficiency of \gaia BBH triples
decreases with increasing metallicity and the decrease is sudden
between $Z=0.005$ and $Z=0.01$. The formation efficiency of BBHs with
orbital periods of $\leq 10^2$ days trends similarly, but the
formation efficiency of all BBHs exhibits a different trend (see
Figure \ref{fig:countBbh}) because the inner BBHs of \gaia BBH triples
have orbital periods of $\lesssim 10^2$ days, as described later
(subsection \ref{sec:InnerBBH}). The number fractions of \gaia BBH
triples to \gaia BH binaries might reflect the number fraction of BBHs
with orbital periods of $\leq 10^2$ days to the sum of isolated BHs
and BHs with non-BH companions: $6$--$9$\% in $Z \le 0.005$, $2$\% in
$Z=0.01$, and $0.2$\% in $Z=0.02$.

\begin{figure}
  \includegraphics[width=\columnwidth]{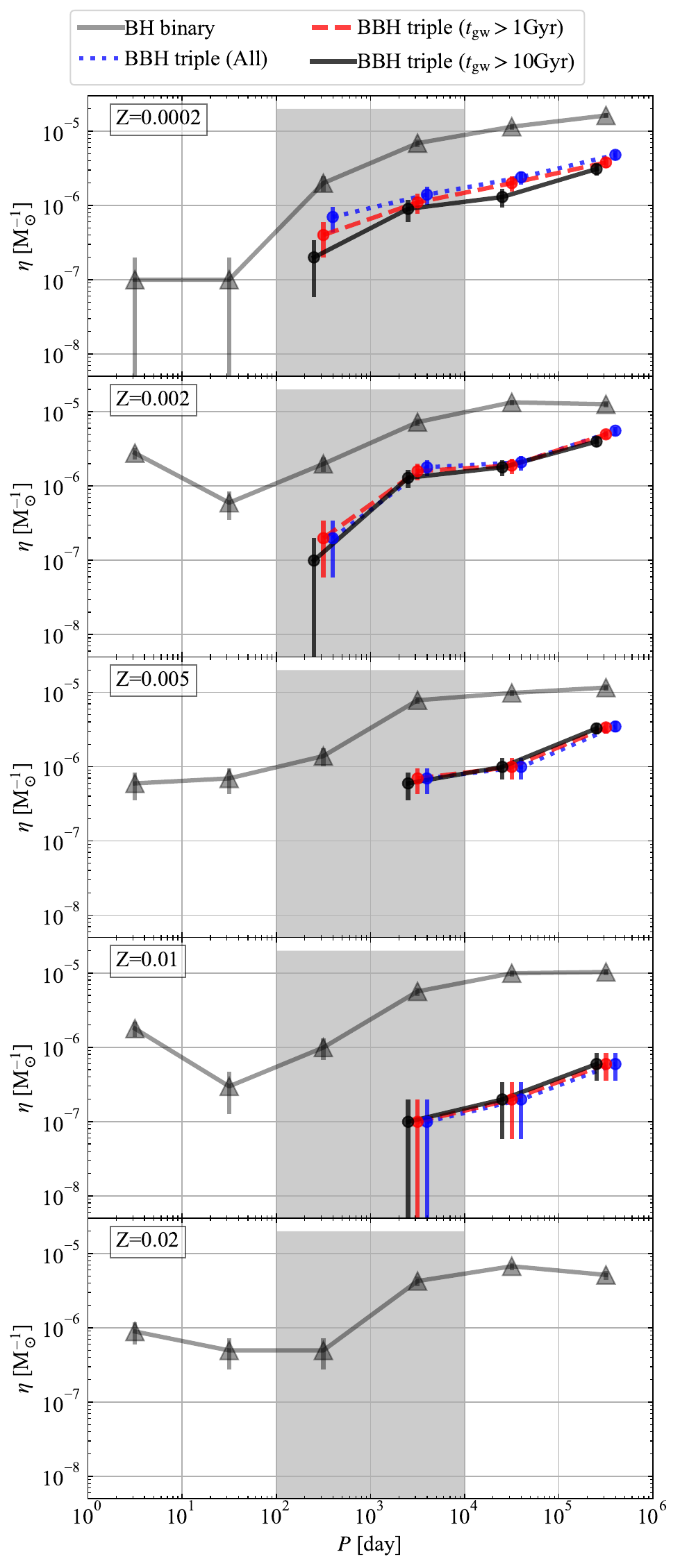}
  \caption{Formation efficiencies of the BH binaries and BBH triples
    in each dex of the orbital period ($\pout$) in different
    metallicity models. Plotted are the formation efficiencies of all
    BBH triples (dotted curves and accompanying circles) and the BBH
    triples with $\tgw>1$ and $\tgw>10$ Gyrs (dashed and solid curves
    with circles, respectively).  The formation efficiency of BH
    binaries is also plotted (solid curve with triangles) as a
    reference.  To improve the visibility, the dotted and solid curves
    with their circles are shifted rightward and leftward by $0.1$
    dex, respectively. Where no circles are plotted, no BBH triples
    were formed in that dex of the orbital period ($\pout$). The error
    bars are described in the caption of Figure
    \ref{fig:etaGaiabbh}. The shaded regions delineate the orbital
    periods ($\pout$) of \gaia BH binaries and \gaia BBH triples.}
  \label{fig:etaGaiabbhOuterperiod}
\end{figure}

As shown in Figure \ref{fig:etaGaiabbhOuterperiod}, the formation
efficiency of the BBH triples reduces with decreasing orbital period
and BBH triples with $P<10^2$ days are not observed. We propose two
reasons for this trend. First, as the orbital period becomes smaller,
BBH triples become dynamically unstable. Second, we observe that the
formation efficiencies of BH binaries also reduce with decreasing
orbital period except at $P<10$ days. A BH (or BBH) capturing a star
tends to be a wide binary (or a wide outer binary), and is ejected
from an open cluster before it evolves to a tight binary. This is
because an open cluster has a low escape velocity.  In some cluster
models ($Z=0.002$, $Z=0.01$ and $Z=0.02$ clusters), the formation
efficiency of BH binaries rises again at $P<10$ days. BH binaries form
not through dynamical capture but through common envelope
evolution. As triple star evolution between BBH and tertiary star is
absent in our cluster model, the formation efficiencies of BBH triples
remain small at $P<10$ day. Even if we replace $\tgw$ with $\tgwacc$
in Figure \ref{fig:etaGaiabbhOuterperiod}, the results are
similar. The reason is the same as in Figure \ref{fig:etaGaiabbh}.

\subsection{Orbits of \gaia BBH triples}
\label{sec:GaiaBBHtriple}

\begin{figure*}
  \includegraphics[width=2\columnwidth]{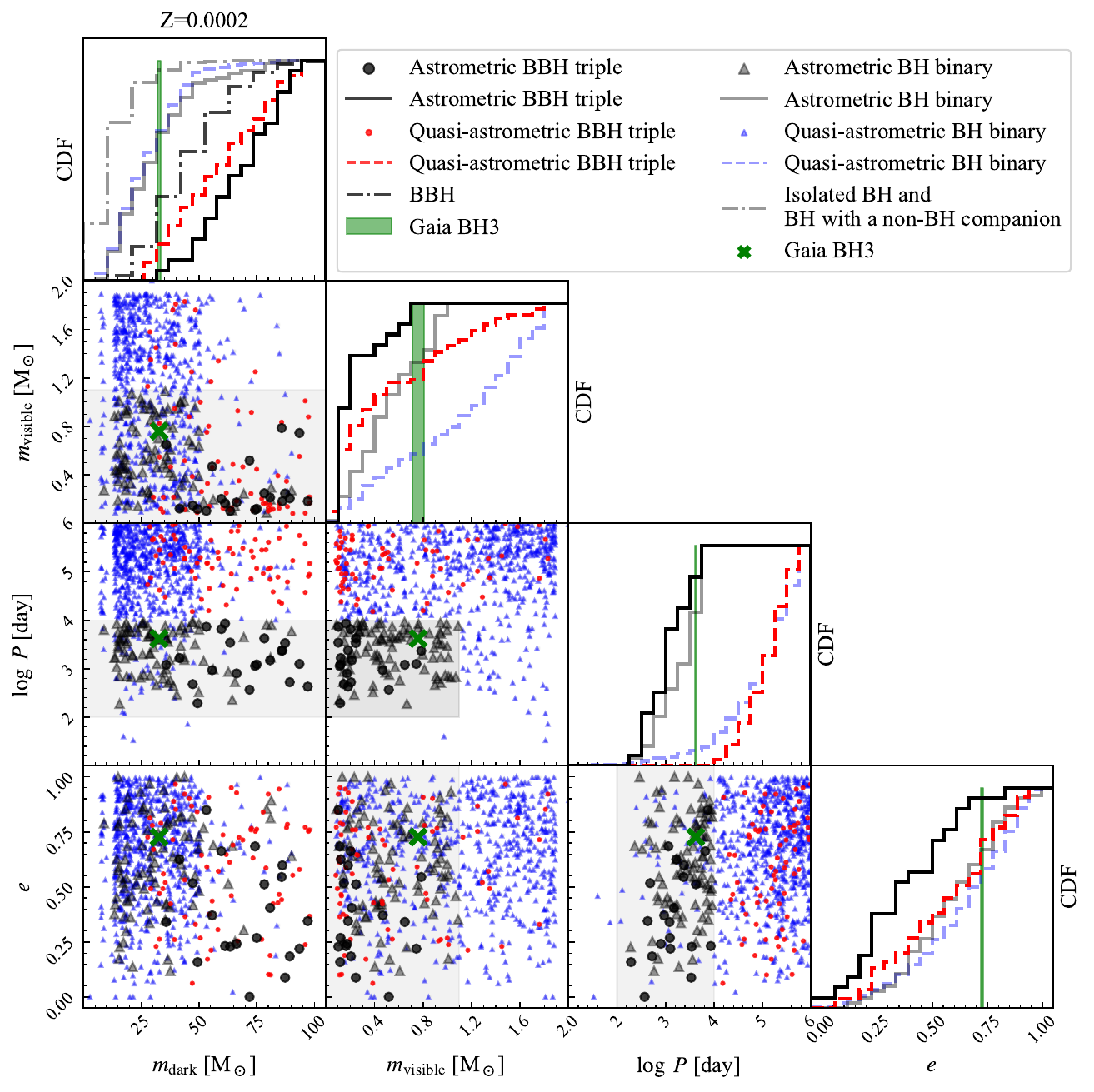}
  \caption{Distributions of dark-object masses ($\mdark$), visible
    star masses ($\mvisi$), orbital periods ($\pout$), and orbital
    eccentricities ($\eout$) of \gaia BBH triples, quasi-\gaia BBH
    triples, \gaia BH binaries, and quasi-\gaia BH binaries in the
    $Z=0.0002$ model. In the cumulative distribution function (CDF)
    plots, the stepwise patterns are the cumulative histograms. The
    top left panel shows the cumulative histograms of the BBHs and
    the summed isolated BHs and BHs with non-BH companions. The
    shaded regions in the other panels delineate the region with
    $\mvisi \le 1.1\;\msun$ and $10^2 \leq \pout/{\rm days} \leq
    10^4$, which supposedly include \gaia BH binaries and \gaia BBH
    triples. The observational parameters of Gaia BH3 are also
    plotted. The CDF plots account for the observational errors of
    Gaia BH3 and the other plots display the typical values of Gaia
    BH3.}
  \label{fig:cornerplotGaiabbh_metl_z2e-4}
\end{figure*}

Figure \ref{fig:cornerplotGaiabbh_metl_z2e-4} shows the distributions
of dark-object masses ($\mdark$), visible star masses ($\mvisi$),
orbital periods ($\pout$), and orbital eccentricities ($\eout$) of
\gaia BBH triples, quasi-\gaia BBH triples, \gaia BH binaries, and
quasi-\gaia BHs in the $Z=0.0002$ model. Similar distributions were
observed in the other models (see Figures
\ref{fig:cornerplotGaiabbh_metl_z2e-3}--\ref{fig:cornerplotGaiabbh_metl_z2e-2}
in Appendix \ref{sec:othermodel}). Therefore, the trends observed in
the $Z=0.0002$ model also hold in the other models unless otherwise
remarked.

\begin{figure}
  \includegraphics[width=\columnwidth]{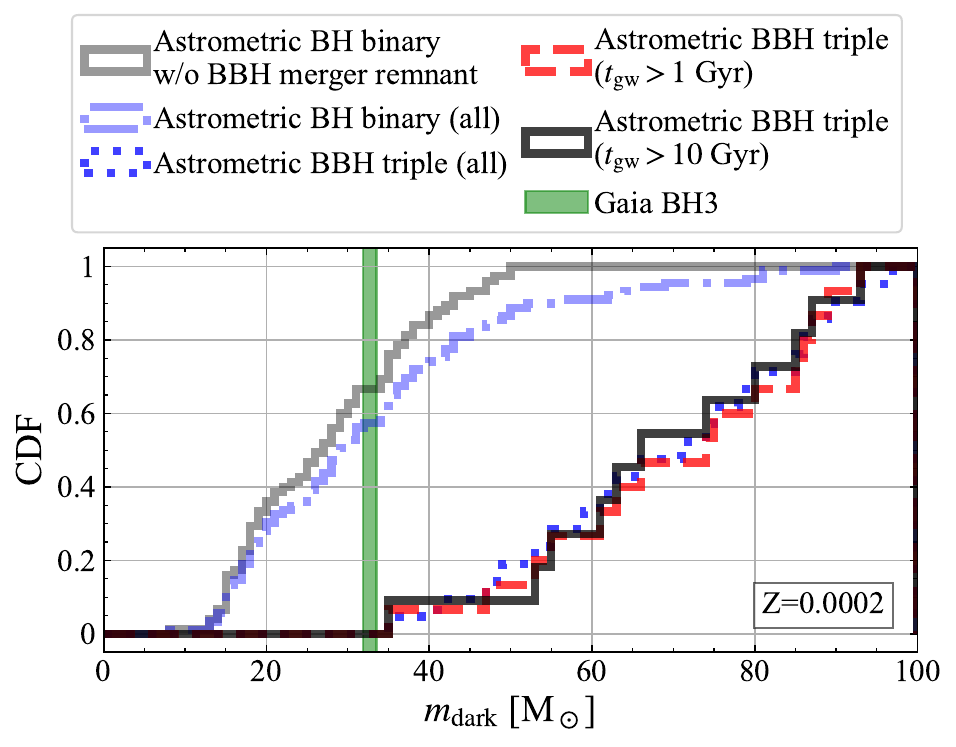}
  \includegraphics[width=\columnwidth]{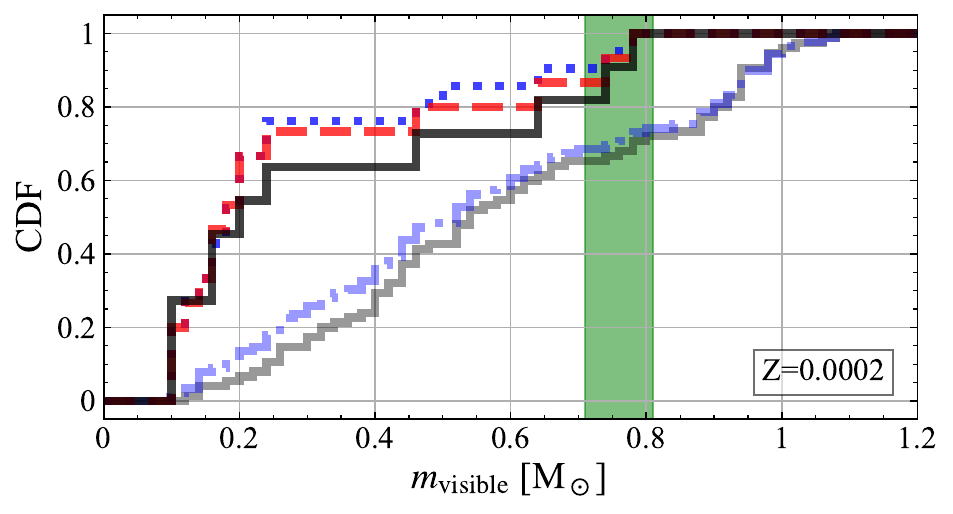}
  \includegraphics[width=\columnwidth]{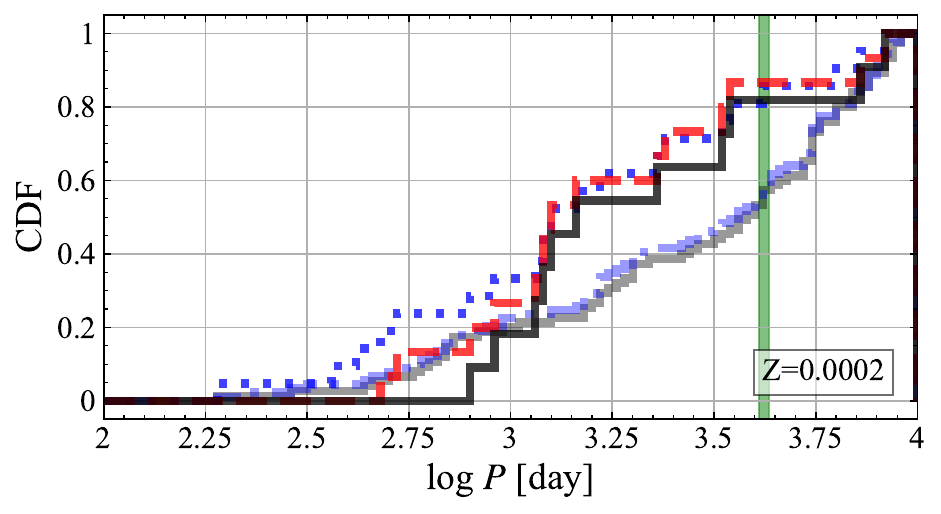}
  \includegraphics[width=\columnwidth]{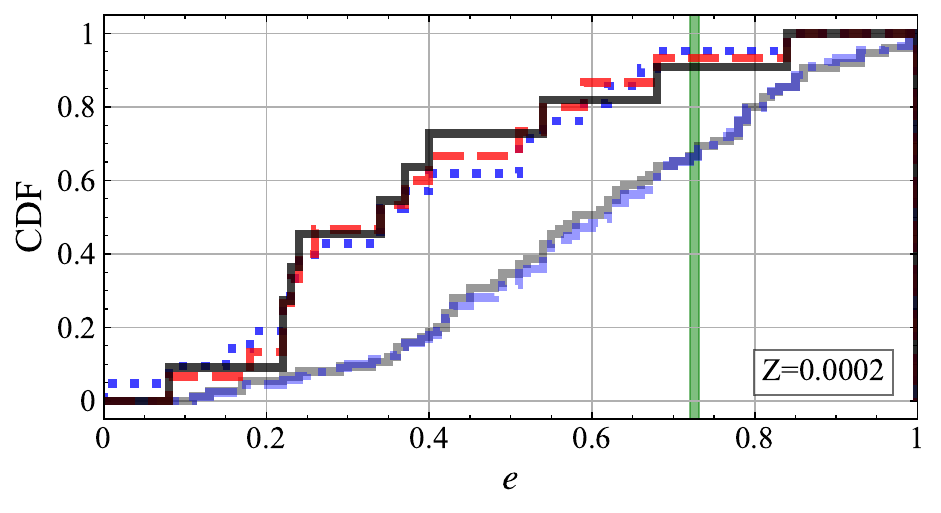}
  \caption{CDFs of dark-object masses ($\mdark$), visible star masses
    ($\mvisi$), orbital periods ($\pout$), and orbital eccentricities
    ($\eout$) of \gaia BBH triples and \gaia BH binaries in the
    $Z=0.0002$ model.  The solid black, dashed, and dotted curves show
    the CDFs of \gaia BBH triples with $\tgw>10$ Gyrs, $\tgw>1$ Gyr,
    and all ages, respectively. The blue dotted-dashed curves include
    all \gaia BH binaries and the solid gray curves exclude \gaia BH
    binaries containing BBH merger remnants. The shaded regions
    delineate the parameter ranges of Gaia BH3.}
  \label{fig:caveats}
\end{figure}

Before discussing these parameters, we first assess whether the
parameters depend on merging and non-merging inner BBHs, as both types
of BBHs are found in \gaia BBH triples. Figure \ref{fig:caveats} shows
the cumulative distributions of these parameters in \gaia BBH triples
with all inner BBHs, $\tgw>1$ Gyrs inner BBHs, and $\tgw>10$ Gyrs
inner BBHs in the $Z=0.0002$ model. The orbital-period distributions
depend on $\tgw$. The number of \gaia BBH triples with $P<10^3$ days
decreases with increasing $\tgw$ but the other parameters are
independent of $\tgw$.

Because the current PETAR code does not implement GW-recoil kicks, our
simulations possibly obtain BH binaries and BBH triples consisting of
BBH merger remnants. These entities are unwanted because the GW-recoil
kicks generated by BBH mergers should eject BBH merger remnants from
their open cluster host, and BBH merger remnants lack opportunities
for capturing visible stars or other BHs. Therefore, we also assessed
whether ignoring the GW-recoil kicks would affect our results. BBH
merger remnants are not found in \gaia BBH and quasi-\gaia BBH
triples, but appear in a portion of non-inner BBHs. The effect of
\gaia BH binaries was strongest in the $Z=0.0002$ cluster model. In
the $Z=0.0002$, $0.002$, $0.005$, $0.01$, and $0.02$ models, BBH
merger remnants appeared in $16$\%, $13$\%, $1.1$\%, $0$\%, and $0$\%
of astrometric BH binaries, respectively. Nevertheless, the formation
efficiencies of \gaia BH binaries decreased only by several $10$\%
after accounting for GW-recoil kicks\footnote{By the same reasoning as
Footnote \ref{note:few10percent}, a decrease of several $10$\% is
deemed insignificant.}. The top panel of Figure \ref{fig:caveats}
compares the BH mass distributions after including and excluding \gaia
BH binaries with BBH merger remnants (blue dotted-dashed and solid
gray curves, respectively) in the $Z=0.0002$ model. When \gaia BH
binaries with BBH merger remnants are excluded, there are no \gaia BH
binaries with $\gtrsim 50\;\msun$ BHs because of pair-instability and
pulsational pair-instability supernovae. Except for this, the results
are identical. Overall, we conclude that GW-recoil kicks can be
ignored except when investigating BH binaries and BBHs with BBH merger
remnants.

As shown in the top left panel of Figure
\ref{fig:cornerplotGaiabbh_metl_z2e-4}, the dark-object masses in the
\gaia BBH triples are double those in the \gaia BH binaries. This
observation is simply explained by the number of BHs in each case: two
in each \gaia BBH triple and one in each \gaia BH binary. The \gaia BH
binaries contain more massive BHs than isolated BHs and BHs with
non-BH companions. Similarly, \gaia BBH triples contain more massive
BBHs than BBHs without tertiary stars because more massive BHs and
BBHs capture visible stars more easily. In addition, as isolated BHs
become less massive, they are more easily ejected from clusters by
natal kicks and unable to capture visible stars.

\begin{figure}
  \includegraphics[width=\columnwidth]{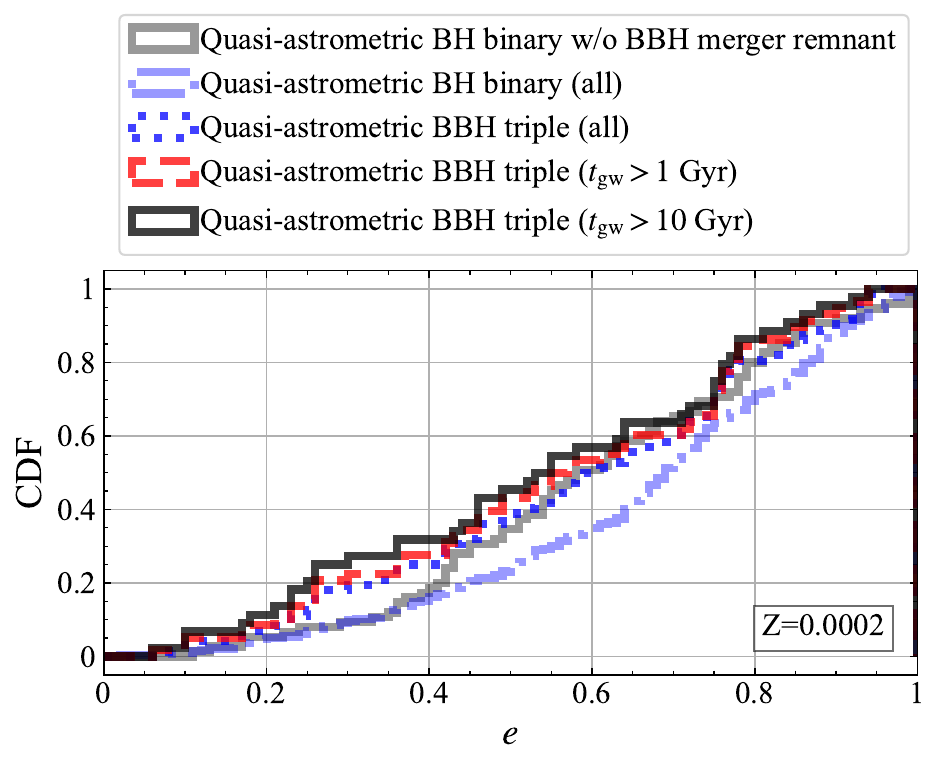}
  \caption{CDFs of the quasi-\gaia BH binaries and quasi-\gaia BBH
    triples (see the bottom panel of Figure \ref{fig:caveats} for
    description)}
  \label{fig:eccNongaiabbh}
\end{figure}

In the BBH triples, the orbital eccentricities decrease with
decreasing orbital period (Figure
\ref{fig:cornerplotGaiabbh_metl_z2e-4}) because BBH triples with small
orbital periods and large orbital eccentricities are unstable. As
confirmation, we compared the eccentricity distributions of \gaia BH
binaries, quasi-\gaia BH binaries, \gaia BBH triples, and quasi-\gaia
BBH triples. As shown in Figure \ref{fig:caveats} (bottom panel) and
Figure \ref{fig:eccNongaiabbh}, the eccentricity distributions of the
BBH triples and BH binaries converge with increasing orbital period of
the BBH triples.

Interestingly, this correlation emerges in the range of orbital
periods ($10^2 \le P/{\rm day} \le 10^4$) suitable for astrometric
observations. In \gaia BBH triples with $P \leq 10^3$ days, the
orbital eccentricities extend to $\sim 0.8$ (see Figure
\ref{fig:cornerplotGaiabbh_metl_z2e-4}), but \gaia BBH triples with $P
\leq 10^3$ days and eccentricities of $\gtrsim 0.8$ are not
expected. This absence might substantially affect the detectability of
\gaia BBH triples (see section \ref{sec:Distinguishability}).

\begin{figure}
  \includegraphics[width=\columnwidth]{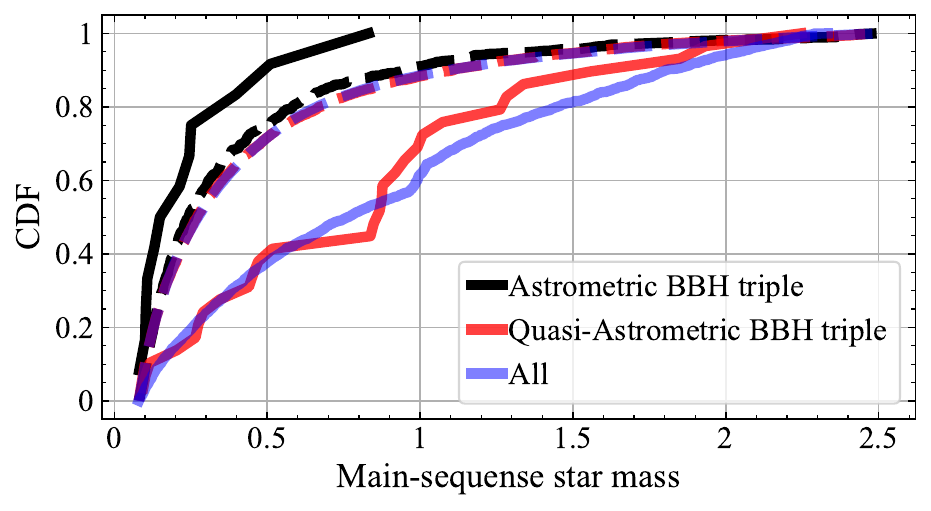}
  \caption{Cumulative mass distributions of $\leq 2.5\;\msun$
    main-sequence stars in the inner parts of clusters forming \gaia
    BBH triples, those forming quasi-\gaia BBH triples, and all
    clusters at $0.04$ Gyrs (dashed curves) and $0.6$ Gyrs (solid
    curves) in the $Z=0.0002$ model. An inner part is defined as a
    region within $1$ pc from the cluster center. As stars with $>
    2.5\;\msun$ evolve off the main sequence at $0.6$ Gyrs, only the
    mass distributions of $\leq 2.5\;\msun$ main-sequence stars are
    shown.}
  \label{fig:massdistincluster}
\end{figure}

As revealed in Figure \ref{fig:cornerplotGaiabbh_metl_z2e-4}, the
visible star masses increase with increasing orbital periods of BBH
triples. This positive correlation is explained by the ejection of
massive main-sequence stars from the inner BBHs of \gaia BBH
triples. In fact, clusters forming \gaia BBH triples contain less
massive visible stars at $0.6$ Gyrs than other clusters (see Figure
\ref{fig:massdistincluster}), although all clusters exhibit similar
mass distributions of visible stars at $0.04$ Gyrs. By the same
reasoning, \gaia BBH triples contain less massive visible stars than
\gaia BH binaries.  If the mass distributions of main-sequence stars
were similar in all clusters, this correlation would result from the
Kozai--Lidov mechanism. However, owing to the activity of inner BBHs
in \gaia BBH triples, the mass distributions of main-sequence stars
intrinsically differ between \gaia-BBH-triple forming clusters and
other clusters.

Next, we compared the dark object mass, visible star mass, and
eccentricity of Gaia BH3 with those of \gaia BH binaries and \gaia BBH
triples obtained in our cluster simulations (see Figures
\ref{fig:cornerplotGaiabbh_metl_z2e-4} and \ref{fig:caveats}). This
comparison was performed in the lowest-metallicity model ($Z=0.0002$)
because the [{\rm Fe/H}] ratio in the visible star of Gaia BH3 is
$-2.76$.  Overall, Gaia BH3 will likely be an \gaia BH binary in the
$Z=0.0002$ model. \Gaia BBH triples should be more massive in
dark-object masses, and less massive in visible star masses. The
eccentricity of Gaia BH3 ($0.726$) lies on the borderline of
acceptability for \gaia BBH triples (see Figure
\ref{fig:cornerplotGaiabbh_metl_z2e-4}) but is deemed acceptable given
the long orbital period ($4190$ days) of \gaia BBH triples.

We remark that the maximum mass of BHs ($\sim 50 \msun$) depends on
the chosen pair instability and pulsational pair-instability
supernovae models. The recently discovered GW event GW190521
\citep{2020PhRvL.125j1102A, 2020ApJ...900L..13A} shows a large
uncertainty in the lower edge of the pair-instability mass gap
\citep[e.g.][]{2018ApJ...863..153T, 2020ApJ...902L..36F,
  2021MNRAS.501L..49K, 2021MNRAS.504..146V, 2021MNRAS.505.2170T,
  2022MNRAS.516.1072C, 2024MNRAS.531.2786K}. Any discovered \gaia BH
binary with a BH mass of $> 50 \msun$ will likely be an actual \gaia
BBH triple, but this assessment strongly depends on the pair
instability and pulsational pair-instability supernovae models. Thus,
to confirm whether a candidate is a \gaia BBH triple, we must find its
radial-velocity modulation, as \cite{2024PASP..136a4202N} searched
for.

\subsection{Inner BBHs in \gaia BBH triples}
\label{sec:InnerBBH}

\begin{figure*}
  \includegraphics[width=1.9\columnwidth]{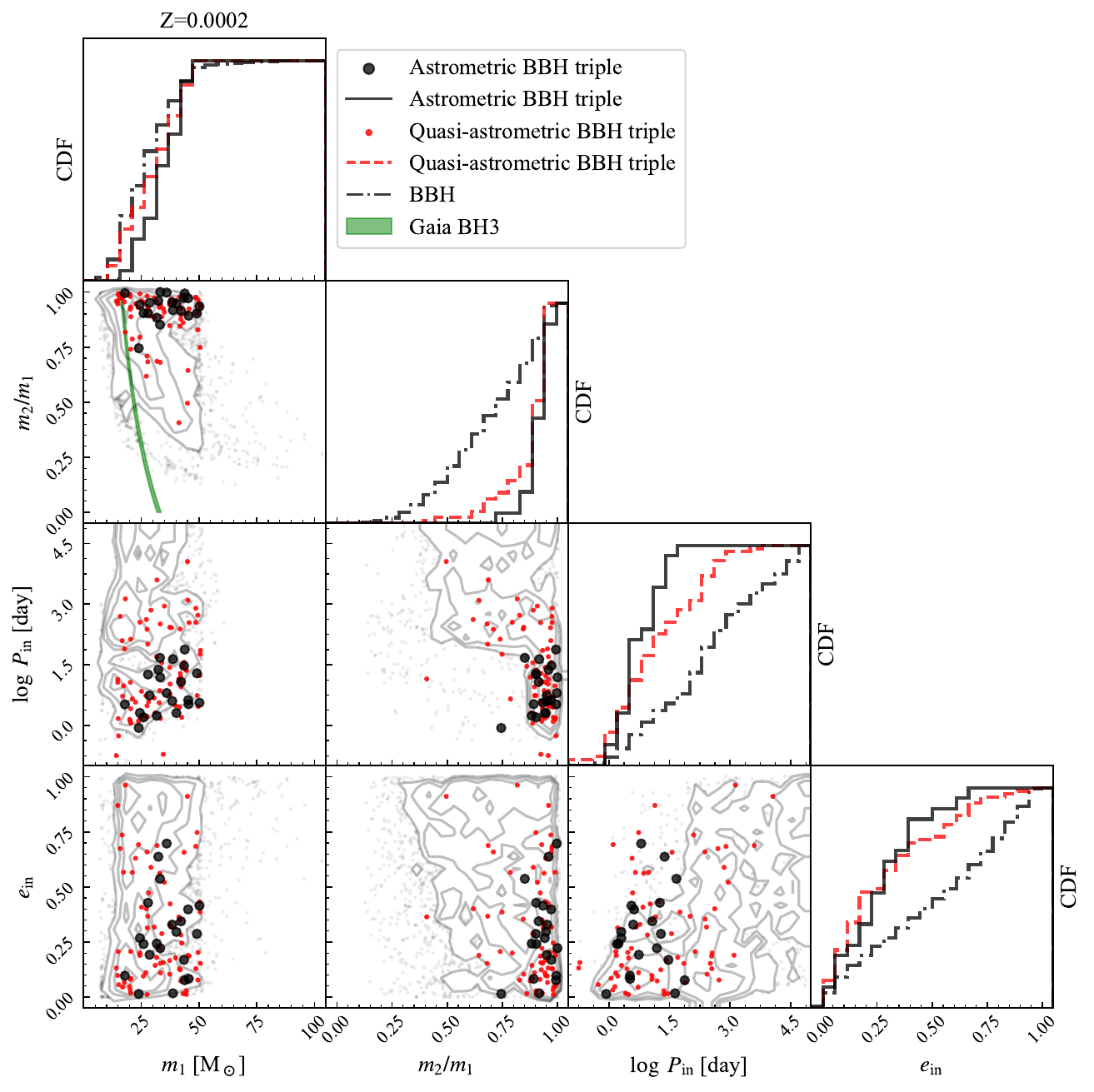}
  \caption{Distributions of primary BH masses ($\mpri$), BH mass
    ratios ($\msec/\mpri$), orbital periods ($\pin$), and orbital
    eccentricities ($\ein$) of the inner BBHs of \gaia BBH triples,
    quasi-\gaia BBH triples, and all BBHs in the $Z=0.0002$ model. The
    CDF plots show the cumulative histograms. In the other panels, the
    distributions of all BBHs are indicated by contours in dense
    regions and by isolated points in the diffuse regions. The plot of
    mass ratio versus primary BH mass also displays the possible
    parameter space of Gaia BH3.}
  \label{fig:cornerplotAllbbh_metl_z2e-4}
\end{figure*}

Figure \ref{fig:cornerplotAllbbh_metl_z2e-4} shows the distributions
of primary BH masses ($\mpri$), BH mass ratios ($\msec/\mpri$),
orbital periods ($\pin$), and orbital eccentricities ($\ein$) of the
inner BBHs in \gaia BBH triples, the inner BBHs of quasi-\gaia BBH
triples, and all BBHs in the $Z=0.0002$ model. The results of the
other models are shown in Figures
\ref{fig:cornerplotAllbbh_metl_z2e-3} --
\ref{fig:cornerplotAllbbh_metl_z2e-2} of Appendix
\ref{sec:othermodel}. In all models except the $Z=0.02$ model, the
inner BBHs of \gaia BBH triples exhibit similar trends but the BH mass
increases with decreasing metallicities.

The inner BBHs of \gaia BBH triples possess orbital periods of
$\lesssim 10^2$ days and mass ratios close to unity (see the scatter
plot of orbital period versus mass ratio in Figure
\ref{fig:cornerplotAllbbh_metl_z2e-4}). Thus, the formation
efficiencies of \gaia BBH triples are strongly correlated with those
of BBHs with orbital periods of $\leq 10^2$ days (see Figure
\ref{fig:countBbh}). The reason for their short orbital periods is
that \gaia BBH triples should be unstable if their inner BBHs have
long orbital periods.

The observed mass ratios are interpreted as follows. Similarly to the
inner BBHs of \gaia BBH triples, all BBHs with orbital periods of
$\lesssim 10^2$ days tend to exhibit mass ratios of $\sim 1$. In other
words, most BBHs that can potentially become inner BBHs of \gaia BBH
triples have near-unity mass ratios.  Quasi-\gaia BBH triples can be
interpreted similarly. The inner BBHs of quasi-\gaia BBH triples
possess mass ratios of $\sim 0.5$--$1$ and their orbital periods
should be limited to $\lesssim 10^3$ days by configurational
stability. In all BBHs with orbital periods of $\lesssim 10^3$ days,
the mass ratios are $\sim0.5$--$1$.

In BBHs with short orbital periods, the mass ratios tend to approach
unity because short-period BH progenitors undergo binary interactions
such as stable mass transfer and common envelope evolution, which
reduce the mass difference between the two BH progenitors. Many binary
population synthesis models have demonstrated near-unity mass ratios
in BBHs with short orbital periods \citep[e.g.][]{2018MNRAS.474.2959G,
  2020A&A...640L..20B, 2022ApJ...926...83T}, consistent with the GW
observations \citep{2023PhRvX..13a1048A}. If there existed many
short-period BBHs with mass ratios of $\lesssim 0.3$
\citep{2022MNRAS.511.1362T}, the deficit of inner BBHs with low mass
ratios would explain the octupole-level interactions in triple systems
\citep{2013MNRAS.431.2155N, 2015MNRAS.452.3610A}. However, such BBHs
are intrinsically a minority in our model.

The simulations yielded BBHs with $\gtrsim 50\;\msun$ BHs, which
should not be formed in reality\footnote{Star clusters can form
$\gtrsim 50\;\msun$ BH through merger of main-sequence star and giant
star \citep{2020MNRAS.497.1043D, 2021MNRAS.507.3612R}. On the other
hand, we do not find $\gtrsim 50\;\msun$ BHs formed through such
mechanism. This is because our cluster models have much smaller mass
density ($20\;\dens$) at the initial time than theirs ($\sim 8 \times
10^3\;\dens$ for $10^3\;\msun$ clusters).}. As described in subsection
\ref{sec:GaiaBBHtriple}, these anomalies form because our simulation
ignores GW-recoil kicks. Nevertheless, their number is small and no
BBH triples with BBH merger remnants are formed. The presence of
unrealistic BBHs does not affect our results.

Inner BBHs of \gaia BBH triples have smaller orbital eccentricities
than all the BBHs. This is also true for inner BBHs of quasi-\gaia BBH
triples. These facts are attributable to two phenomena. First, these
BBHs tend to have smaller orbital periods than all the BBHs, and thus
eccentric inner BBHs of \gaia BBH triples and quasi-\gaia BBH triples
merge due to orbital decay of GW radiation before our simulations end.
Second, the eccentric inner BBHs of \gaia BBH triples and quasi-\gaia
BBH triples should be dynamically unstable.

\begin{figure}
  \includegraphics[width=\columnwidth]{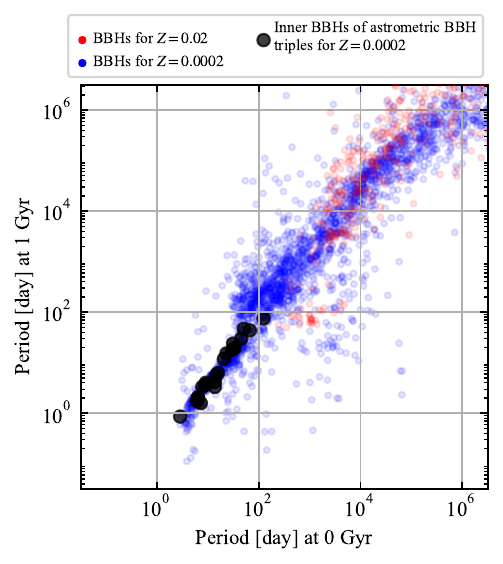}
  \caption{Scatter plot of final orbital periods of the BBHs in the
    $Z=0.02$ and $Z=0.0002$ models and the inner BBHs of \gaia BBH
    triples in the $Z=0.0002$ model versus initial orbital periods of
    their progenitors.}
  \label{fig:initfinperiod}
\end{figure}

The lack of \gaia BBH triples in the $Z=0.02$ model (as described in
the previous subsections) can be attributed to the small number of
BBHs with orbital periods of $\lesssim 10^2$ days (see Figures
\ref{fig:countBbh} and
\ref{fig:cornerplotAllbbh_metl_z2e-2}). Similarly, no quasi-\gaia BBH
triples were formed because few BBHs possess orbital periods of
$\lesssim 10^3$ days. To explain the absence of \gaia BBH triples, we
note that in the $Z=0.02$ model (which simulates a solar-metallicity
environment), primordial binaries in close orbits do not enter into
the common envelope evolution, but merge when either of their members
fills the Roche lobe. This merging coincides with the time of
Hertzsprung gap phases in solar-metallicity stars. Hertzsprung-gap
stars are supposed not to enter into common envelope evolution,
because the contrast at the boundaries between their helium cores and
hydrogen envelopes may not be steep
\citep[e.g.][]{2004ApJ...601.1058I, 2007ApJ...662..504B,
  2012ApJ...759...52D, 2015ApJ...814...58D, 2017MNRAS.472.2422M,
  2018MNRAS.474.2959G, 2018MNRAS.480.2011G, 2022MNRAS.516.5737B,
  2022ApJ...926...83T, 2023MNRAS.524..426I}. Note that this argument
holds in our binary evolution model. Depending on the choice of binary
evolution model, the formation efficiency of \gaia BBH triples may
increase in the $Z=0.02$ model because the number of short-period BBHs
increases in this model. \cite{2023MNRAS.524..426I} presented the
formation efficiencies of merging BBHs in various binary evolution
models, such as the different $\lambda_{\rm CE}$ models of
\cite{2010ApJ...716..114X} and
\cite{2021A&A...645A..54K}. \cite{2022MNRAS.516.5737B} showed that a
small Wolf--Rayet wind factor can relax the sudden decrease of BBH
formation efficiency at $Z \gtrsim 0.02$.

This argument is corroborated by Figure \ref{fig:initfinperiod}. In
the $Z=0.0002$ model, massive main-sequence binaries with orbital
periods of $\lesssim 10^2$ days evolve into BBHs with orbital periods
of $\lesssim 10^2$ days. A portion of these evolved BBHs become the
inner BBHs of \gaia BBH triples. In contrast, the corresponding
main-sequence binaries in the for $Z=0.02$ model can hardly evolve to
BBHs with orbital periods of $\lesssim 10^2$ days, as rationalized
above.

The above assumption that Hertzsprung-gap stars cannot enter into the
common envelope evolution is largely uncertain. In an alternative
binary evolution model, Hertzsprung-gap stars can enter into common
envelope evolution and the formation efficiency of short-period BBHs
drastically increases, even at $Z=0.02$
\citep[e.g.][]{2020A&A...636A.104B}. Thus, the formation efficiencies
of \gaia BBH triples may be comparable in the $Z=0.02$ and $Z=0.0002$
models.  The effects of this uncertainty, estimated in the alternative
binary evolution model, are presented in Appendix
\ref{sec:alternative}.

\section{Detectability of astrometric BBH triples}
\label{sec:Distinguishability}

In this section, we detect \gaia BBH triples among the \gaia BH binary
candidates based on precise radial-velocity measurements of tertiary
stars (subsection \ref{sec:spectroscopy}) and space-borne GW
observatories (subsection \ref{sec:gravitationalwave}).

\subsection{Short-time radial-velocity modulations of tertiary stars}
\label{sec:spectroscopy}

\begin{figure*}
  \includegraphics[width=2\columnwidth]{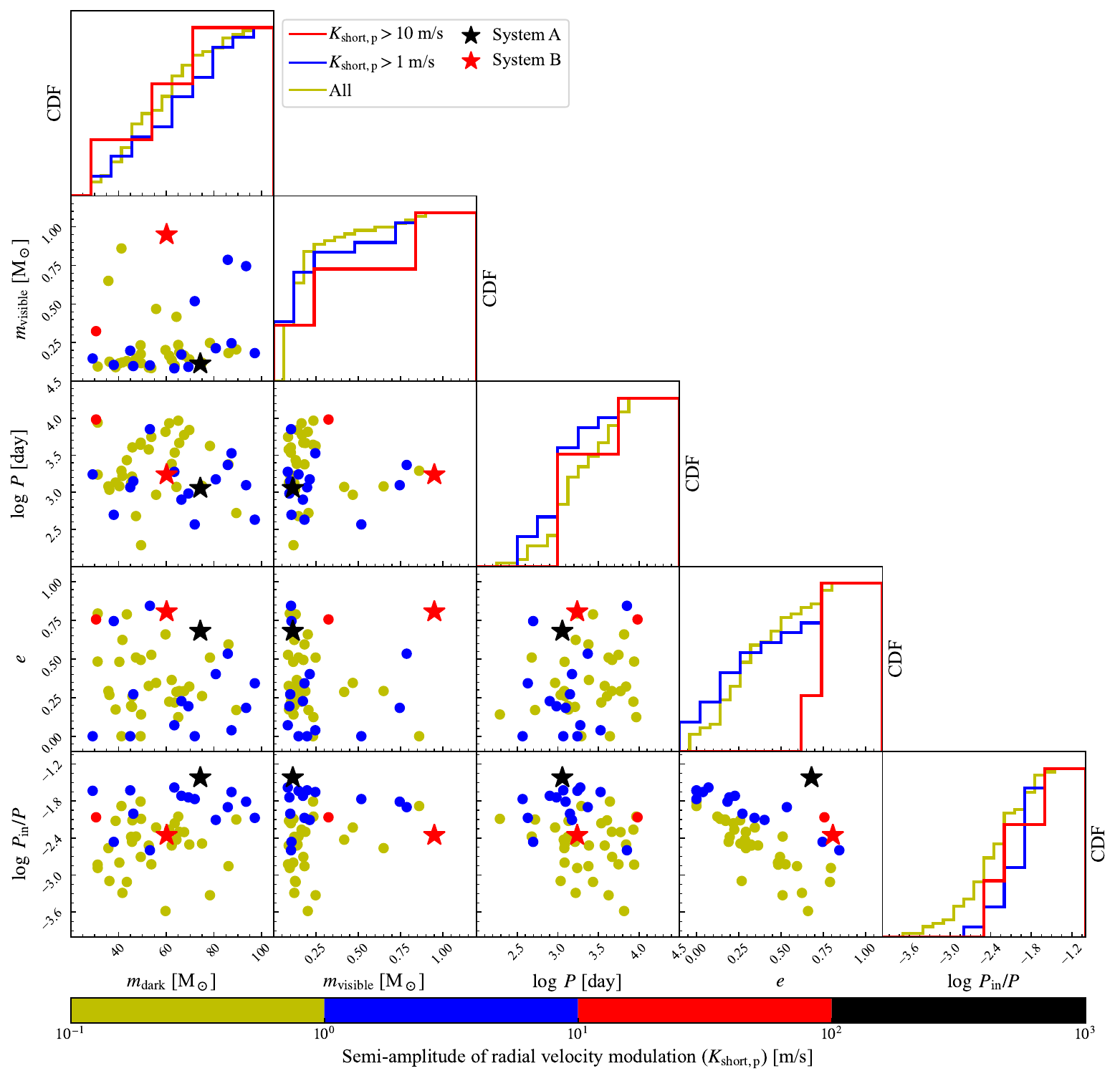}
  \caption{Radial-velocity modulations ($\kshortp$) as functions of
    \gaia BBH triple parameters: dark-object mass ($\mdark$), visible
    star mass ($\mvisi$), outer period ($\pout$), outer eccentricity
    ($\eout$), and ratio of inner-to-outer periods ($\pin/\pout$). The
    outer orbits are viewed edge-on ($I_{\rm obs}=90^\circ$). The CDFs
    of these parameters are plotted for $\kshortp>10$ m/s,
    $\kshortp>1$ m/s, and all $\kshortp$. The black and red star
    symbols indicate systems A and B, respectively (see Figure
    \ref{fig:rvmodulation} for details).}
  \label{fig:cornerouter}
\end{figure*}

The inner BBH perturbs the motion of a tertiary star from its Kepler
orbit. As first proposed by \cite{2020ApJ...890..112H} and
\cite{2020ApJ...897...29H}, this deviation is directly detectable as
short-term radial-velocity modulation, which depends on the orbital
configuration of the triples and is maximized at the pericenter
passage of the tertiary star.

The semi-amplitude of the radial-velocity modulation is approximated as
\begin{align}
  \kshortp = K_{\rm short} (1-\eout)^{-7/2} \label{eq:kshortp}
\end{align}
\citep{2023ApJ...958...26H}. Here, $K_{\rm short}$ denotes the
semi-amplitude of the radial-velocity modulation of a circular orbit
($e=0$), calculated as \citep{2008A&A...491..899M}:
\begin{align}
  K_{\rm short} = &\frac{\mpri \msec}{m_{12}^2} \left(
  \frac{m_{123}}{m_{12}} \right)^{-2/3} \left( \frac{\pin}{\pout}
  \right)^{7/3} \nonumber \\
  &\times \left( \frac{2\pi Gm_{12}^3}{m_{123}^2 \pout} \right)^{1/3}
  \sin I_{\rm obs}
\end{align}
where $m_{12}=\mpri+\msec$, $m_{123}=\mpri+\msec+\mvisi$, and $I_{\rm
  obs}$ is the inclination of the outer orbit relative to the
observer.

In Figure \ref{fig:cornerouter}, the semi-amplitudes of the
radial-velocity modulations at the pericenter passages ($\kshortp$) of
all $48$ \gaia BBH triples formed in our simulations are projected
onto their system parameter planes. For simplicity, we assume that the
outer orbit is viewed edge-on ($I_{\rm obs}=90^\circ$). Among the $48$
systems, we identified $17$, $3$, and $1$ triples with $\kshortp$
values above $1$, $10$, and $100$ m/s, respectively. At the precision
of the current high-resolution spectroscopy, which is much better than
$\sim 1$ m/s, these triples are detectable from radial-velocity
measurements in principle, although their detectability depends on the
distances, stellar types, and inclinations of their systems.

To quantitatively examine the detectability, we performed 3-body
simulations of the simulated triples with $>100$ m/s $\kshortp$
(system A) and $>10$ m/s $\kshortp$ (system B), yielding more precise
radial velocities than the analytic approximations
(\ref{eq:kshortp}). For this purpose, we adopted a fast and accurate
direct $N$-body integrator called TSUNAMI \citep[see][for
  details]{2023IAUS..362..404T}.

\begin{figure*}
  \centering
  \includegraphics[width=1.8\columnwidth]{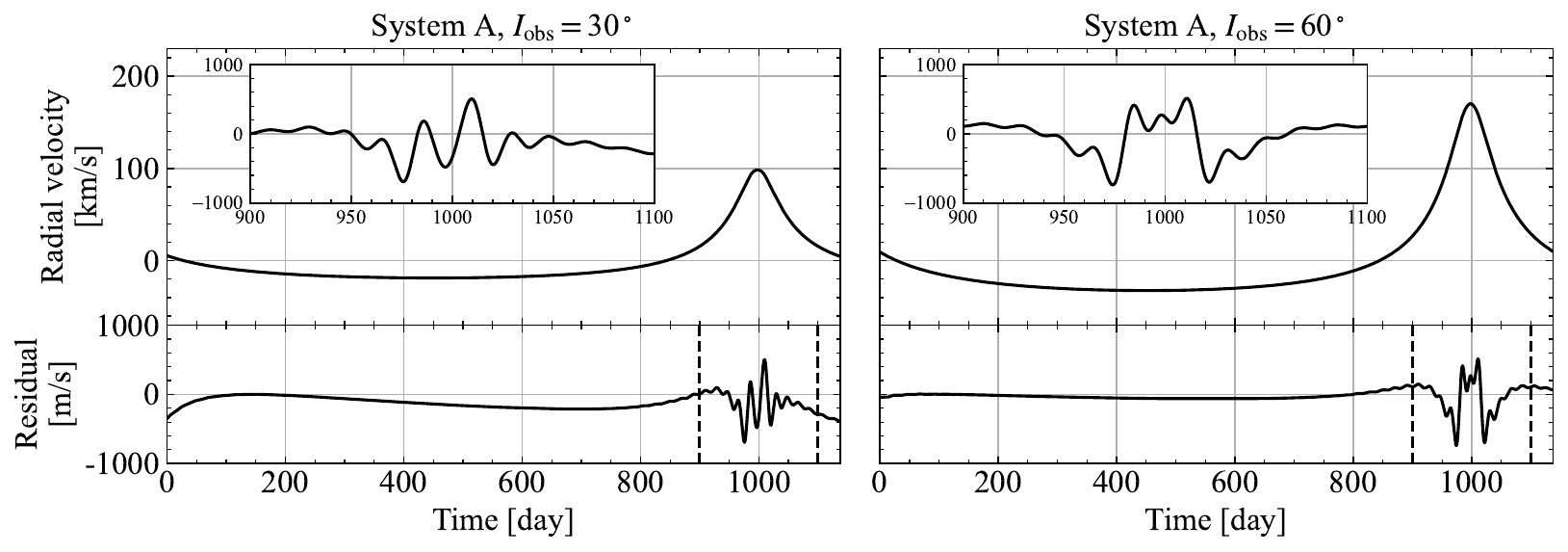}
  \includegraphics[width=1.8\columnwidth]{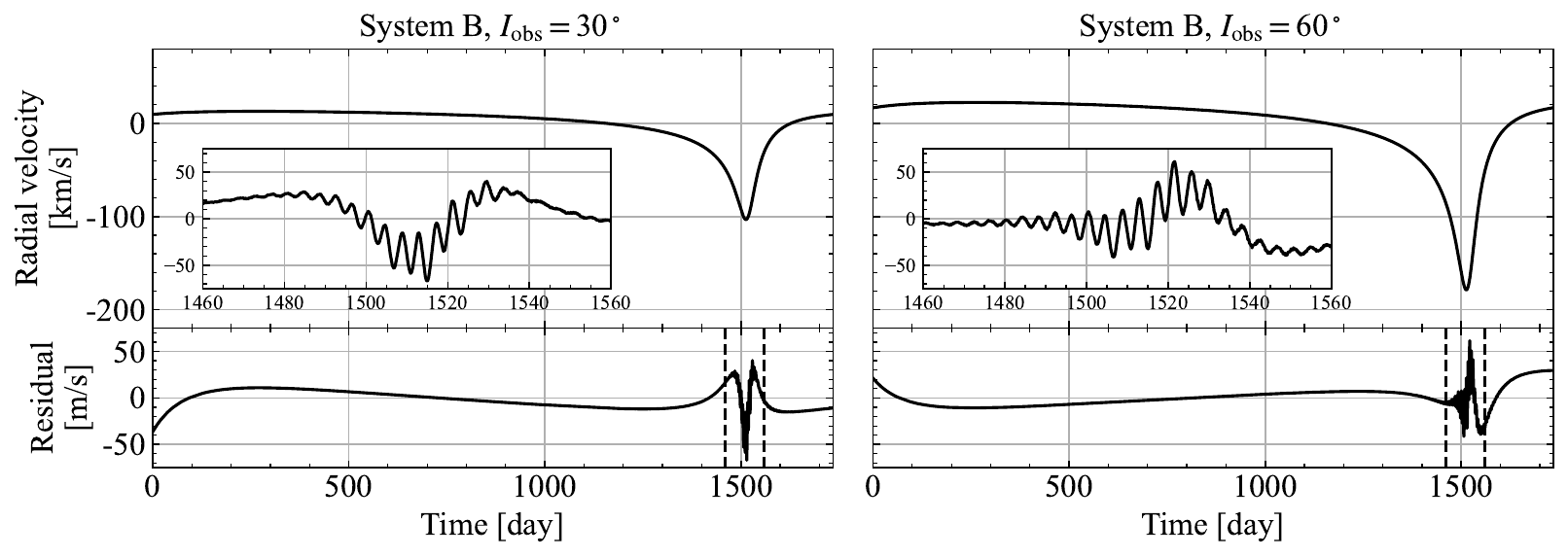}
  \caption{Temporal evolutions of radial velocities and their
    residuals from the Kepler motions for systems A and B in Figure
    \ref{fig:cornerouter}, observed at inclinations of $30^\circ$
    (left) and $60^\circ$ (right) from the outer orbits.}
  \label{fig:rvmodulation}
\end{figure*}

Figure \ref{fig:rvmodulation} shows the radial-velocity curves of the
two triples over one outer-orbital period. The residuals of the Kepler
orbital velocities were extracted using the open-source
radial-velocity fitting toolkit RadVel \citep[see][for
  details]{2018PASP..130d4504F}. The velocity modulations of both
triples exhibit spike-like structures with semi-amplitudes of $\sim
100$ m/s around the pericenter passage, which last for more than one
month. The period of the modulations is approximately $\pin/2$,
clarifying the presence of an inner binary companion.

Such radial-velocity modulations are enlarged in \gaia BBH triples
with larger outer-orbital eccentricities and/or larger ratios of
inner-to-outer-orbital periods (see the $\pin/\pout$ versus $e$ plot
of Figure \ref{fig:cornerouter}). Because the period ratios of \gaia
BBH triples are not easily determined in advance, \gaia BH binaries
with larger (outer) eccentricities are assumed as promising candidates
for \gaia BBH triples. In fact, $\kshortp$ exceeded $1$ m/s in more
than half of our simulated \gaia BBH triples with $e>0.7$.

With an eccentricity of $0.726$, Gaia BH3 is a promising \gaia BBH
triple candidate, but the masses of its dark object and visible star
($\mdark=33\;\msun$ and $\mvisi=0.8\;\msun$, respectively) differ from
those of our simulated samples. Nevertheless, it is worth performing
high-resolution spectroscopy for Gaia BH3 around at its pericenter
passage, the next of which is August 23rd 2029.

\subsection{Space-borne GW observatory}
\label{sec:gravitationalwave}

\begin{figure}
  \includegraphics[width=\columnwidth]{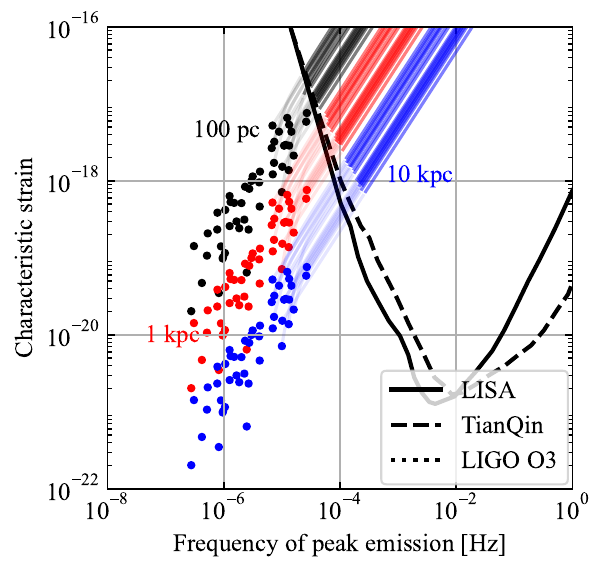}
  \caption{Characteristic strains at the peak emission frequency of
    the Inner BBHs of \gaia BBH triples at various distances
    ($100$ pc, $1$ kpc, and $10$ kpc). Dots indicate the inner
    BBHs at a cluster age of $1$ Gyrs and the solid lines trace the
    evolutions of merging of inner BBHs. GWs from the inner BBHs of
    \gaia BBH triples at $100$ pc, $1$ kpc, and $10$ kpc can be
    detected when $\tgw<100$, $<10$, and $<1$ Myrs, respectively
    (thick lines). The calculations are described in
    \cite{2019PhRvD..99f3003K}, \cite{2022MNRAS.515.5106W}, and
    \cite{2024MNRAS.533.2262L}. The sensitivity curves of LISA (solid
    black curve) and TianQin (dashed black curve) are derived from the
    data of \cite{2019CQGra..36j5011R} and \cite{2019PhRvD.100d3003W},
    respectively.}
  \label{fig:gravitationalwave}
\end{figure}

We now assesses whether GW emissions from the inner BBHs of \gaia BBH
triples can be detected by space-borne GW observatories such as the
Laser Interferometer Space Antenna
\citep[LISA:][]{2017arXiv170200786A} and TianQin
\citep{2016CQGra..33c5010L, 2025RPPh...88e6901L}. An \gaia BBH triple
can be distinguished from an \gaia BH binary if the inner BBH of the
\gaia BBH triple emits detectable GWs. Figure
\ref{fig:gravitationalwave} shows the characteristic strains at the
peak emission frequencies of the inner BBHs at distances of $100$ pc,
$1$ kpc, and $10$ kpc.  We can see that GWs from non-merging inner
BBHs cannot be detected, and that GWs from merging inner BBHs at $100$
pc, $1$ kpc, and $10$ kpc can be detected when their $\tgw < 100$,
$10$, and $1$ Myrs, respectively.

We estimate that $< 0.1$ \% of inner BBHs of \gaia BBH triples can be
observed by GWs for the following reason. \Gaia BBH triples in our
simulations are metal-poor origins, and the members of the galactic
halo components. The observable number increases proportionally to
$D^3$, where $D$ is the distance to an \gaia BBH triple. In contrast,
the timescale of detectable GWs from each inner BBH is proportional to
$D^{-1}$. Therefore, the number of inner BBHs of \gaia BBH triples
that can be detected in GW observations increases proportionally to
$D^2$. More distant \gaia BBH triples will more likely contain inner
BBHs emitting detectable GWs.  Given the distance of Gaia BH2 ({$1.2$
  kpc.}), we expect that \gaia BBH triples can be detected by Gaia
astrometry and its followup spectroscopy at distances up to $\sim 1$
kpc. GWs from these \gaia BBH triples can be detected $10$ Myrs before
their mergers, as described above. Approximately $0.1$\% of the
merging inner BBHs of \gaia BBH triples can exist in this phase during
$10$ Gyrs. Moreover, not all \gaia BBH triples contain merging inner
BBHs. Ultimately, $< 0.1$\% of the inner BBHs of \gaia BBH triples
should be observable in GWs. To detect GWs from the inner BBHs of
\gaia BBH triples, we require $10^3$ \gaia BBH triples and $10^4$
\gaia BH binaries formed in metal-poor environments within $1$
kpc. Therefore, space-borne GW observatories are unlikely to detect
GWs from the inner BBHs of \gaia BBH triples.

In the above estimate, we assumed $\sim 1$ kpc as the detection
horizon of \gaia BH binaries. \cite{2025PASP..137d4202N} predicted
that in Gaia Data Release 4 (DR4) and Final Data Release (Final DR),
the detection horizon of \gaia BH binaries will extend to $\sim 10$
kpc. The GWs of the inner BBH of a \gaia BBH triple located at $10$
kpc can be detected only $1$ Myrs before the merger of the inner
BBH. The existence probability of such an inner BBH is very small
($0.01$\%), although Gaia DR4 and Final DR are expected to discover
around $30$ and $45$ \gaia BH binaries, respectively
\citep{2025PASP..137d4202N}. The expected number of \gaia BBH triples
detectable by space-borne GW observatories is much smaller than $1$.

\section{Discussion and Conclusion}
\label{sec:DiscussionAndConclusion}

We investigated the intrinsic probability of \gaia BBH triples, or
BBHs hidden within \gaia BH binary candidates. The existence
probability is high in low-metallicity environments ($\sim10$\% and
$\sim1$\% at $Z \le 0.005$ and $0.01$, respectively) and zero in
solar-metallicity environments ($Z=0.02$). This result is reflected by
the formation efficiency of BBHs with orbital periods of $\lesssim
10^2$ days, which also depends on the adopted binary evolution model.
Such BBHs can capture visible stars as tertiary stars, forming triple
systems (\gaia BBH triples) with outer-orbital periods of
$10^2$--$10^4$ days. Half of the inner BBHs of \gaia BBH triples are
non-merging BBHs and the remainder are usually characterized by long
GW radiation on timescales of $\tgw > 1$ Gyrs. We also confirmed that
3-body effects do not accelerate the merger processes of the inner
BBHs in \gaia BBH triples.

\Gaia BBH triples exhibited three distinct features. First, the
dark-object masses in \gaia BBH triples are twice those of \gaia BH
binaries, simply because \gaia BBH triples contain two BHs. Second,
the visible star mass is lower in \gaia BBH triples than in \gaia BH
binaries. This feature arises not from 3-body effects but from the
intrinsic mass distribution of the clusters forming \gaia BBH
triples. During cluster evolution, the inner BBHs of \gaia BBH triples
eject massive visible stars through binary-single and binary-binary
interactions. Third, \gaia BBH triples move in more circular orbits
than \gaia BH binaries because eccentric \gaia BBH triples are
unstable.

The mass ratios of the inner BBHs of \gaia BBH triples is close to
unity, reflecting the near-unity mass ratios in BBHs with orbital
periods of $\lesssim 10^2$ days. Such BBH progenitors experience
binary interactions that bring BH progenitor masses closer. The orbits
are more circular in the inner BBHs than in all BBHs because they are
influenced by tertiary stars; moreover, the eccentric inner BBHs have
already merged through GW radiation.

If the radial-velocity precision in the spectroscopic observations is
$\sim1$ m/s, then radial-velocity modulations should distinguish
$30$\% of our \gaia BBH triples from \gaia BH binary candidates. \Gaia
BBH triples are more easily distinguished from \gaia BH binaries when
their orbital eccentricities are large and their ratios of
inner-to-outer BBH period are high, thereby enlarging the
radial-velocity modulations. The detection probability of GWs from the
inner BBH of an \gaia BBH is $<0.1$\%.

Based on our simulation results (see Figures
\ref{fig:cornerplotGaiabbh_metl_z2e-4} and \ref{fig:caveats}), we
inferred that Gaia BH3 is probably not an \gaia BBH triple. In
particular, its dark-object mass is too small and its visible star is
too massive. Its eccentricity also appears excessively high in the
bottom panel of Figure \ref{fig:caveats}), but is within the
acceptable limits given the long orbital period ($\sim4190$ days) of
Gaia BH3. If Gaia BH3 is a genuine \gaia BBH triple, its high orbital
eccentricity ($0.726$) possibly gives rise to large radial-velocity
modulations around its pericenter passage.

As a guideline for discovering \gaia BBH triples among a large number
of \gaia BH binaries, we suggest that the following \gaia BH binaries
are \gaia BBH triple candidates that are easily distinguishable from
\gaia BH binary candidates:
\begin{itemize}
\item \Gaia BH binary candidates with low metallicity ($Z \le
  0.005$) visible stars. \Gaia BBH triples with solar metallicities
  have not been found.
\item \Gaia BH binary candidates with twice the expected
  theoretical BH mass. \Gaia BBH triples possess two BHs.
\item \Gaia BH binary candidates with half the expected mass of
  visible stars. The inner BBHs of \gaia BBH triples tend to eject
  high-mass stars from their open-cluster host, providing little
  opportunity for high-mass star capture.
\item \Gaia BH binary candidates with high eccentricities ($\gtrsim
  0.7$) and long periods ($\gtrsim 1000$ days). If they are \gaia BBH
  triples, they have large radial velocity modulations at their
  pericenter passages. The periods should be $\gtrsim 1000$ days,
  otherwise such eccentric \gaia BBH triples are unstable. Such
    modulations will be strong indicators of \gaia BBH triples.
\end{itemize}

As no \gaia BBH triples were found in the $Z=0.02$ model (see Figure
\ref{fig:etaGaiabbh}), we claim that \gaia BH binaries formed in
low-metallicity environments are likelier candidates of \gaia BBH
triples than \gaia BH binaries in solar-metallicity
environments. However, this assessment may strongly depend on the
adopted binary evolution model. The binary evolution model assumes
that when a star in the Hertzsprung-gap phase fills its Roche lobe and
its mass transfer destabilizes, that the Hertzsprung-gap star merges
with its companion star because it may not establish a steep density
gradient between the helium core and the hydrogen envelope
\citep{2004ApJ...601.1058I}. This assumption largely reduces the
formation efficiency of BBHs with orbital periods of $\lesssim 10^2$
days in solar-metallicity environments \citep{2012ApJ...759...52D,
  2018MNRAS.474.2959G, 2022ApJ...926...83T}. If this assumption is
invalid, many \gaia BBH triples should form in solar-metallicity
environments. The discovery of an \gaia BBH triple in a
solar-metallicity environment would impose strong constraints on BBH
formation scenarios.

According to the above guidelines, an \gaia BH binary candidate can be
an \gaia BBH triple if its BH mass is unexpectedly large. However,
this suggestion depends on the selected single-star evolution and
supernova models. Past discoveries of unexpectedly massive BHs, such
as LB-1 \citep{2019Natur.575..618L}\footnote{The nature of the unseen
object of LB-1 has been disputed in several studies
\citep{2020Natur.580E..11A, 2020MNRAS.493L..22E,
  2021MNRAS.502.3436E}.} and GW190521 \citep{2020PhRvL.125j1102A,
  2020ApJ...900L..13A}, have been followed by modifications of
single-star evolution and supernova models \citep{2018ApJ...863..153T,
  2020ApJ...890..113B, 2020ApJ...902L..36F, 2021MNRAS.501L..49K,
  2021MNRAS.504..146V, 2021MNRAS.505.2170T, 2022MNRAS.516.1072C,
  2024MNRAS.531.2786K}, highlighting the importance of placing
observational constraints on \gaia BBH triples
\citep[e.g.][]{2024PASP..136a4202N}. The observational discovery of an
\gaia BBH triple is expected to influence further developments in
single-star evolution and supernova models.

We estimated the number of \gaia BBH triples formed in open clusters
in the Milky Way galaxy, focusing only on \gaia BBH triples, which can
be distinguished from \gaia BH binaries by high-resolution
spectroscopy. Here, we assume that the Gaia mission can initially
detect \gaia BBH triples as \gaia BH binaries at distances within $1$
kpc similar to the distance to Gaia BH2, the most distant \gaia BH
binary \citep{2023ApJ...946...79T, 2023MNRAS.521.4323E}. The number
was calculated as follows:
\begin{align}
  &N_{\rm BBH,<1kpc} \sim 2.7 \left( \frac{\eta_{\rm BBH}}{10^{-6}
    \msun} \right) \left( \frac{f_{\rm cluster}}{0.1} \right) \left(
  \frac{f_{\rm Z<0.01,<1kpc}}{0.75} \right) \nonumber \\
  &\times \left( \frac{M_{\rm MW,<1kpc}}{1.2 \times 10^8 \msun}
  \right) \left( \frac{f_{\rm Kshort,p>1m/s}}{0.3} \right),
\end{align}
where $\eta_{\rm BBH}$ is the formation efficiency of \gaia BBH
triples in open clusters (see Figure \ref{fig:etaGaiabbh}), $f_{\rm
  cluster}$ is the mass fraction of stars born in open clusters
\citep{2006A&A...459..113M, 2007A&A...468..151P}, $f_{\rm
  Z<0.01,<1kpc}$ is the stellar-mass fraction of $Z<0.01$ bodies
within 1 kpc from the Earth, and $M_{\rm MW,<1kpc}$ is the stellar
mass within 1 Kpc from the Earth. We derived $f_{\rm Z<0.01,<1kpc}$
and $M_{\rm MW,<1kpc}$ using the Milky Way model of
\cite{2022ApJ...937..118W}. As more than $30$\% of \gaia BBH triples
express radial-velocity modulations of $>1$ m/s, we set the
distinguishable fraction ($f_{\rm Kshort,p>1m/s}$) to $0.3$.

As a reference, we also calculated the expected number of \gaia BH
binaries within $1$ kpc in the same Milky Way galaxy model as
\begin{align}
  N_{\rm BH,<1kpc} &\sim 120 \left( \frac{\eta_{\rm BH}}{10^{-5}
    \msun} \right) \left( \frac{f_{\rm cluster}}{0.1} \right)
  \nonumber \\
  &\times \left( \frac{M_{\rm MW,<1kpc}}{1.2 \times 10^8 \msun}
  \right).
\end{align}
Note that the formation efficiency of \gaia BH binaries is almost
constant from low metallicities to the solar metallicity (see Figure
\ref{fig:etaGaiabbh}). As $\sim 1$ \gaia BBH triple can be
distinguished among $40$ \gaia BH binaries, a distinguishable \gaia
BBH triple may be discovered before $40$ \gaia BH binaries are
found. \Gaia BH binaries that are \gaia BBH triples can be more easily
found than actual \gaia BH binaries because they contain more massive
dark objects and produce wider swings of their tertiary visible stars
than \gaia BH binaries with the same orbital period. Thus, Gaia
astrometry can detect \gaia BBH triple candidates more efficiently
than \gaia BH binaries. In other words, $40$ discovered \gaia BH
binaries may contain more than one \gaia BBH
triple. \cite{2025PASP..137d4202N} predicts that approximately $30$
and $45$ \gaia BH binaries will be detected by Gaia DR4 and Final DR,
respectively. In the era of these data releases, we can expect to
discover one or more \gaia BBH triples.

Even the discovery of one \gaia BBH triple will exert important and
complementary impacts. A triple containing a BBH merging within the
Hubble time will constrain the formation mechanism of BBHs observed by
GW observatories \citep{2023PhRvX..13d1039A}, whose origins have been
debated \citep{2023PhRvX..13a1048A}. GW observations can reveal the BH
masses and spins of a BBH. On the other hand, an \gaia BBH triple will
provide its position, velocity, and metallicity via its visible
companion star. Its position and velocity will be helpful to make
clear what galactic components (the Galactic disk, halo, bulge, and
center) the \gaia BBH triple belong to. Note that the sky localization
and distance obtained from GW observations are unhelpful for
elucidating the origin of a BBH, because a huge number of galaxies can
be located at the possible position of the BBH. The metallicity of the
\gaia BBH triple's companion will indicate the metallicity of the
observed BBHs.  Several studies have claimed that lower-metallicity
stars are responsible for more massive BBHs found by GW observatories
\citep[e.g.][]{2016Natur.534..512B, 2018MNRAS.474.2959G,
  2021MNRAS.504L..28K, 2022ApJ...926...83T, 2023MNRAS.524..307S}. The
discovery of an \gaia BBH triple would consolidate these claims. A BBH
in an \gaia BBH triple that remains stable during the Hubble time
would be the first case of a non-merging BBH.

\section*{acknowledgments}

We would like to thank anonymous referees for many helpful comments.
This research could not be accomplished without the support by
Grants-in-Aid for Scientific Research, 24K07040 (A.T.), 23K25908
(Y.S.), and 23KJ1153 (T.H.) from the Japan Society for the Promotion
of Science.  M.F. is supported by The University of Tokyo Excellent
Young Researcher Program. L.W. thanks the support from the
one-hundred-talent project of Sun Yat-sen University, the Fundamental
Research Funds for the Central Universities, Sun Yat-sen University
(22hytd09), the National Natural Science Foundation of China through
grant 21BAA00619 and 12233013 and the High-level Youth Talent Project
(Provincial Financial Allocation) through the grant
2023HYSPT0706. A.A.T acknowledges support from the European Horizon
Europe research and innovation program under the Marie
Sk\l{}odowska-Curie grant agreements no. 101103134. Numerical
simulations are carried out on Small Parallel Computers at Center for
Computational Astrophysics, National Astronomical Observatory of
Japan, the Yukawa Institute Computer Facility, and Cygnus/Pegasus at
the CCS, University of Tsukuba.

\appendix

\section{Results of models other than the $Z=0.0002$ model}
\label{sec:othermodel}

\begin{figure*}
  \includegraphics[width=\columnwidth]{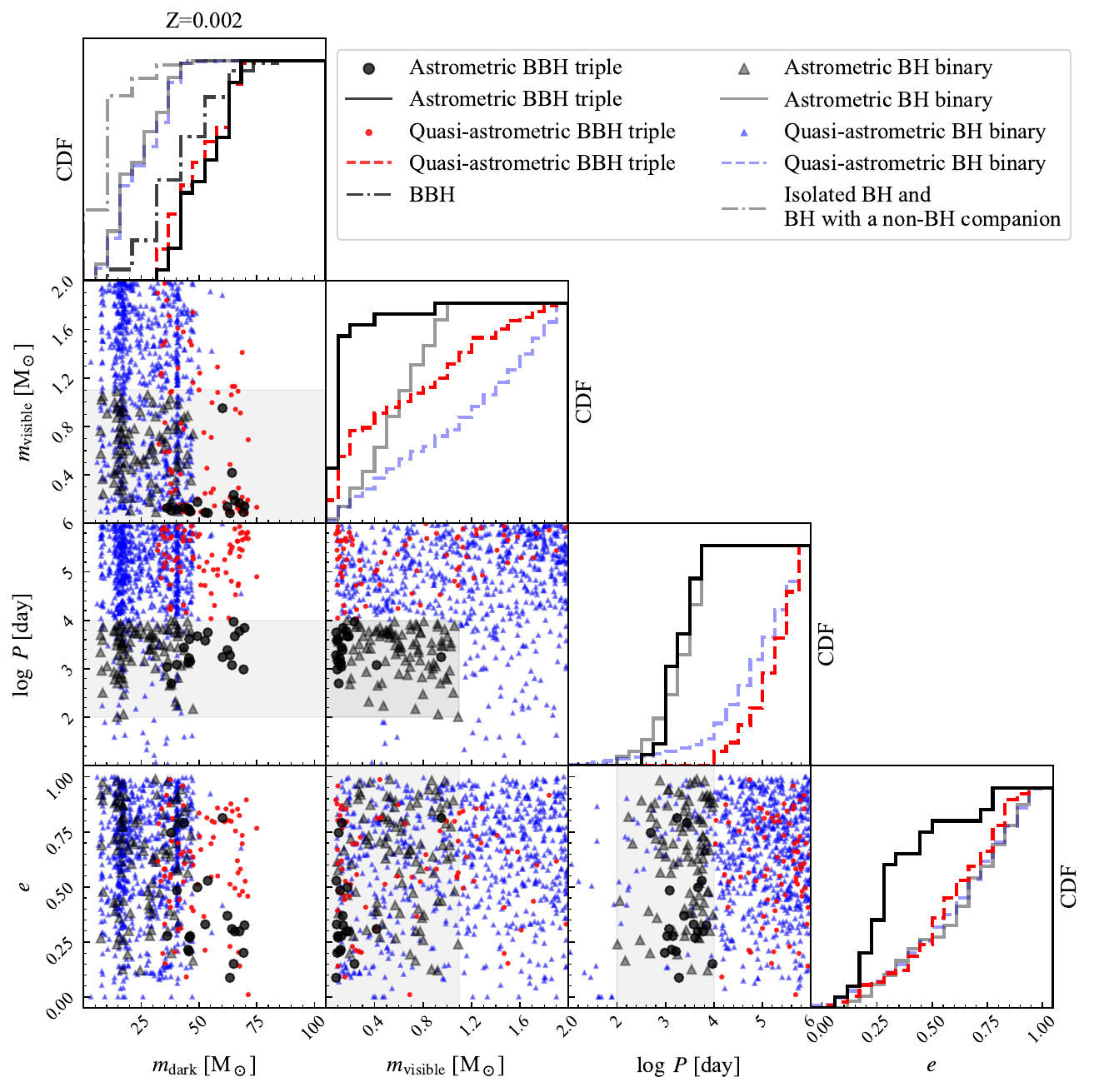}
  \caption{The same as Figure \ref{fig:cornerplotGaiabbh_metl_z2e-4}
    but for the $Z=0.002$ model. No observational parameters of Gaia
    BH3 are plotted because the metallicity in this model far exceeds
    the metallicity of the visible star in Gaia BH3.}
  \label{fig:cornerplotGaiabbh_metl_z2e-3}
\end{figure*}

\begin{figure*}
  \includegraphics[width=\columnwidth]{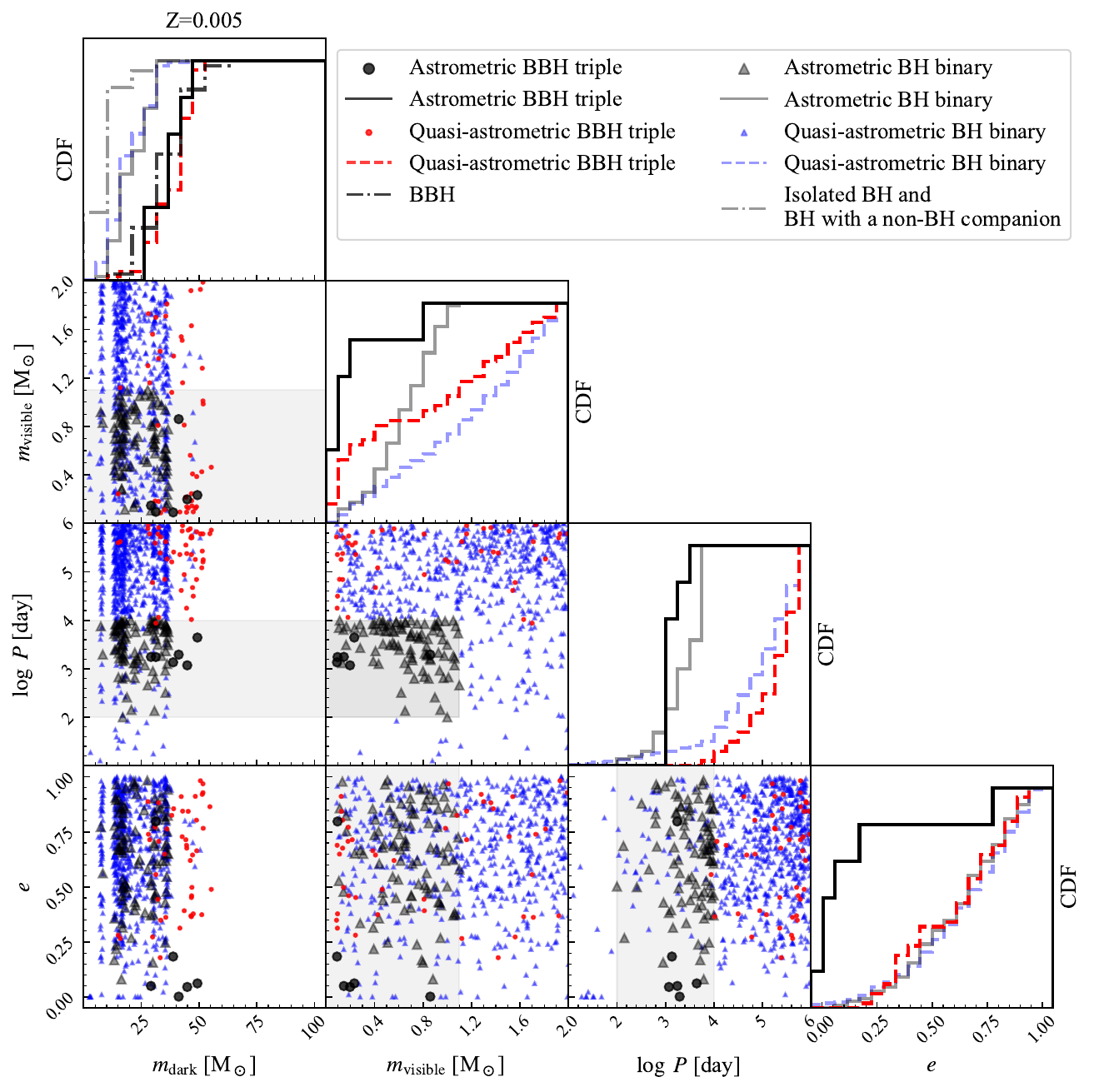}
  \caption{The same as Figure \ref{fig:cornerplotGaiabbh_metl_z2e-3}
    but for the $Z=0.005$ model.}
  \label{fig:cornerplotGaiabbh_metl_z5e-3}
\end{figure*}

\begin{figure*}
  \includegraphics[width=\columnwidth]{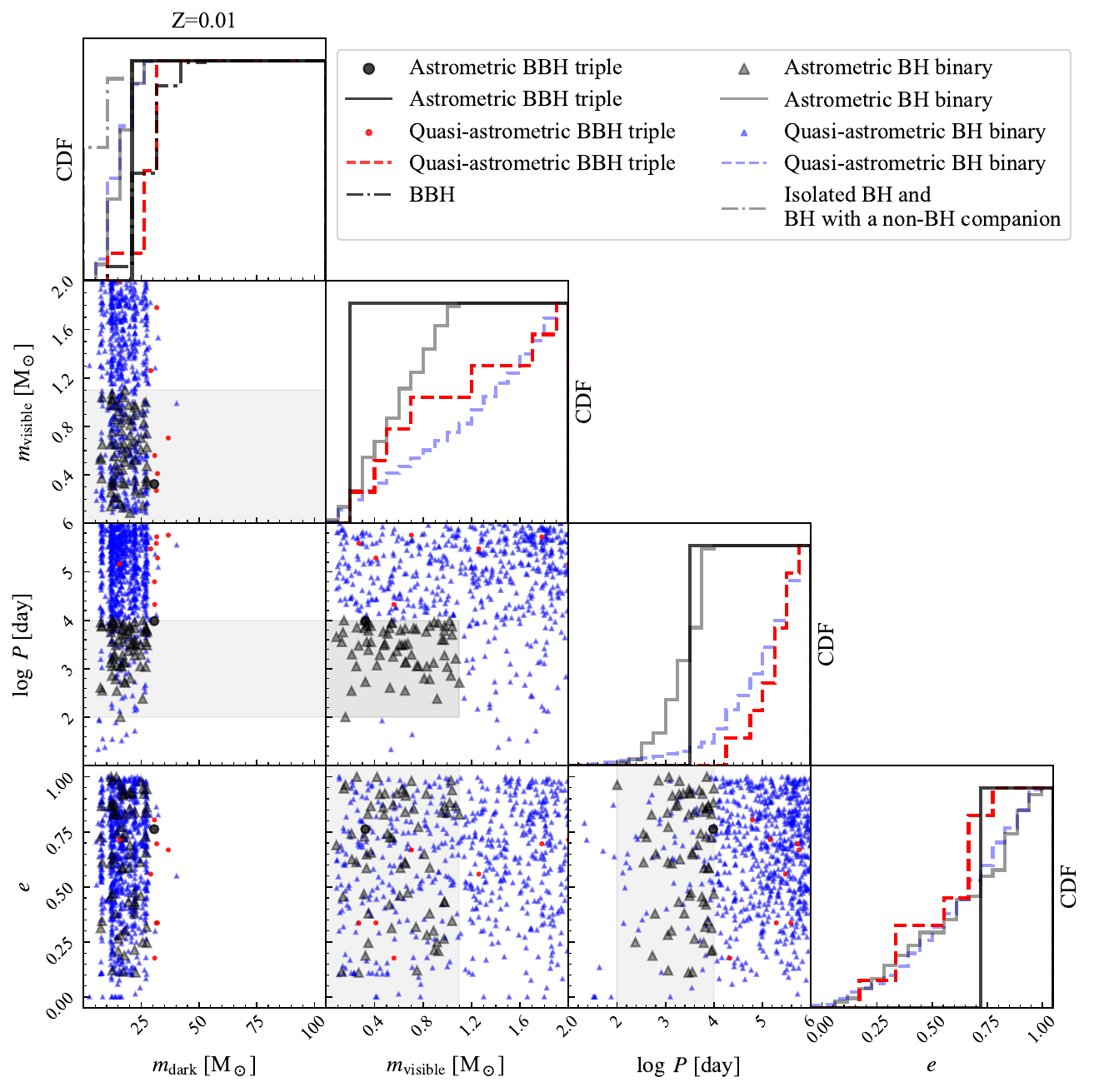}
  \caption{The same as Figure \ref{fig:cornerplotGaiabbh_metl_z2e-3}
    but for the $Z=0.01$ model.}
  \label{fig:cornerplotGaiabbh_metl_z1e-2}
\end{figure*}

\begin{figure*}
  \includegraphics[width=\columnwidth]{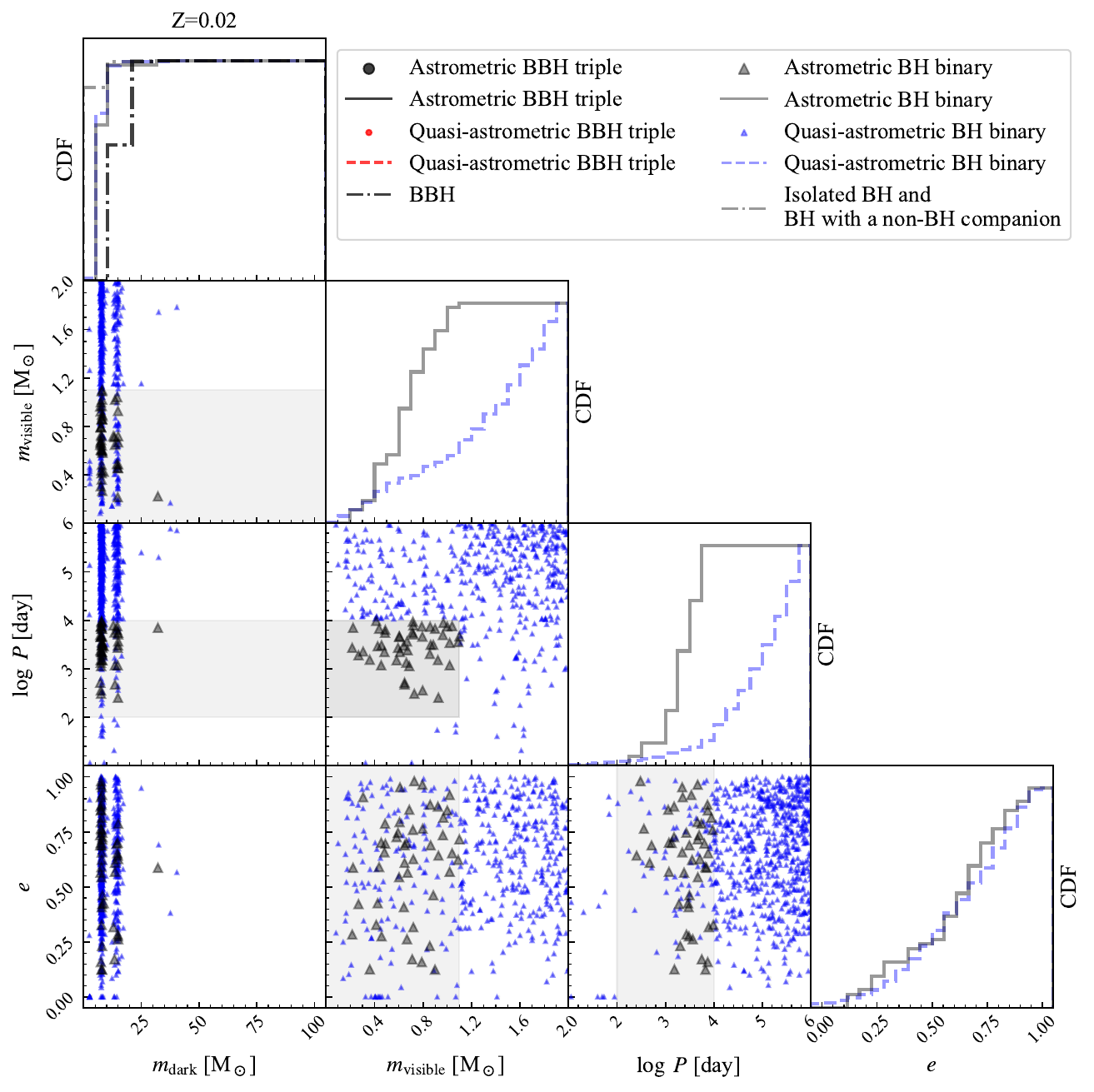}
  \caption{The same as Figure \ref{fig:cornerplotGaiabbh_metl_z2e-3}
    but for the $Z=0.02$ model. Note the absence of \gaia BBH triples
    and quasi-\gaia BBH triples.}
  \label{fig:cornerplotGaiabbh_metl_z2e-2}
\end{figure*}

\begin{figure*}
  \includegraphics[width=\columnwidth]{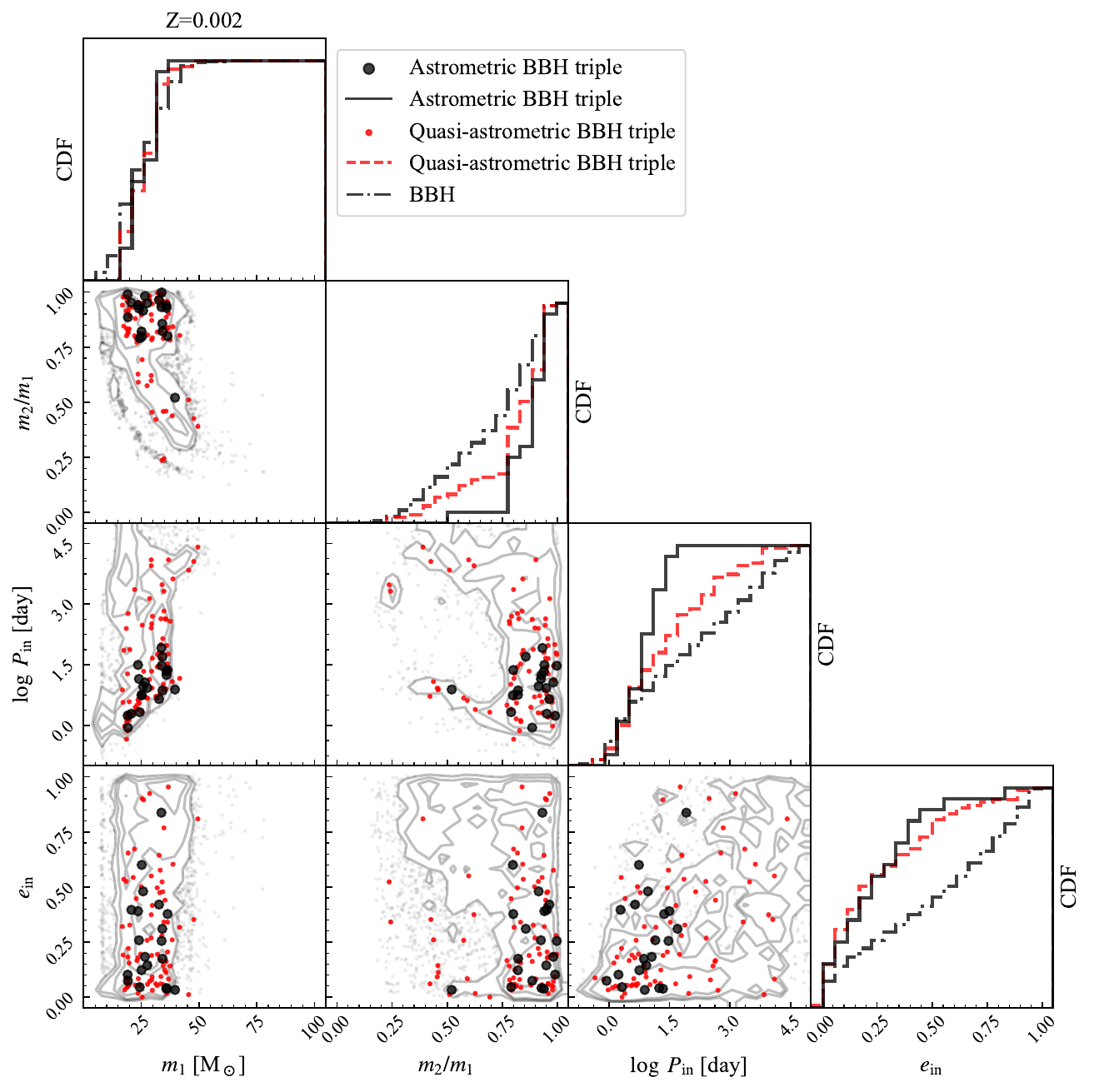}
  \caption{The same as Figure \ref{fig:cornerplotAllbbh_metl_z2e-4} but
    for the $Z=0.002$ model.}
  \label{fig:cornerplotAllbbh_metl_z2e-3}
\end{figure*}

\begin{figure*}
  \includegraphics[width=\columnwidth]{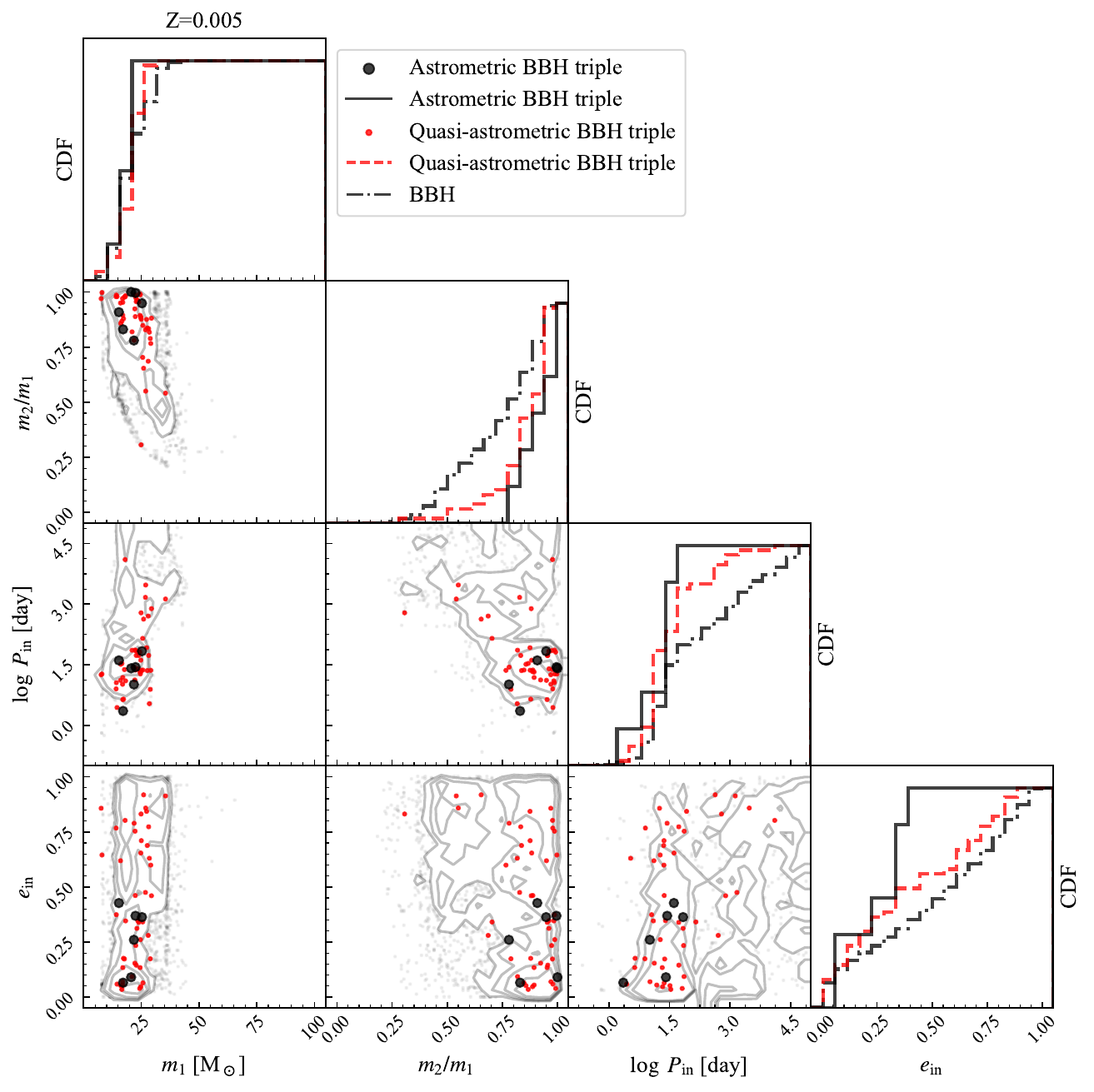}
  \caption{The same as Figure \ref{fig:cornerplotAllbbh_metl_z2e-4}
    but for the $Z=0.005$ model.}
  \label{fig:cornerplotAllbbh_metl_z5e-3}
\end{figure*}

\begin{figure*}
  \includegraphics[width=\columnwidth]{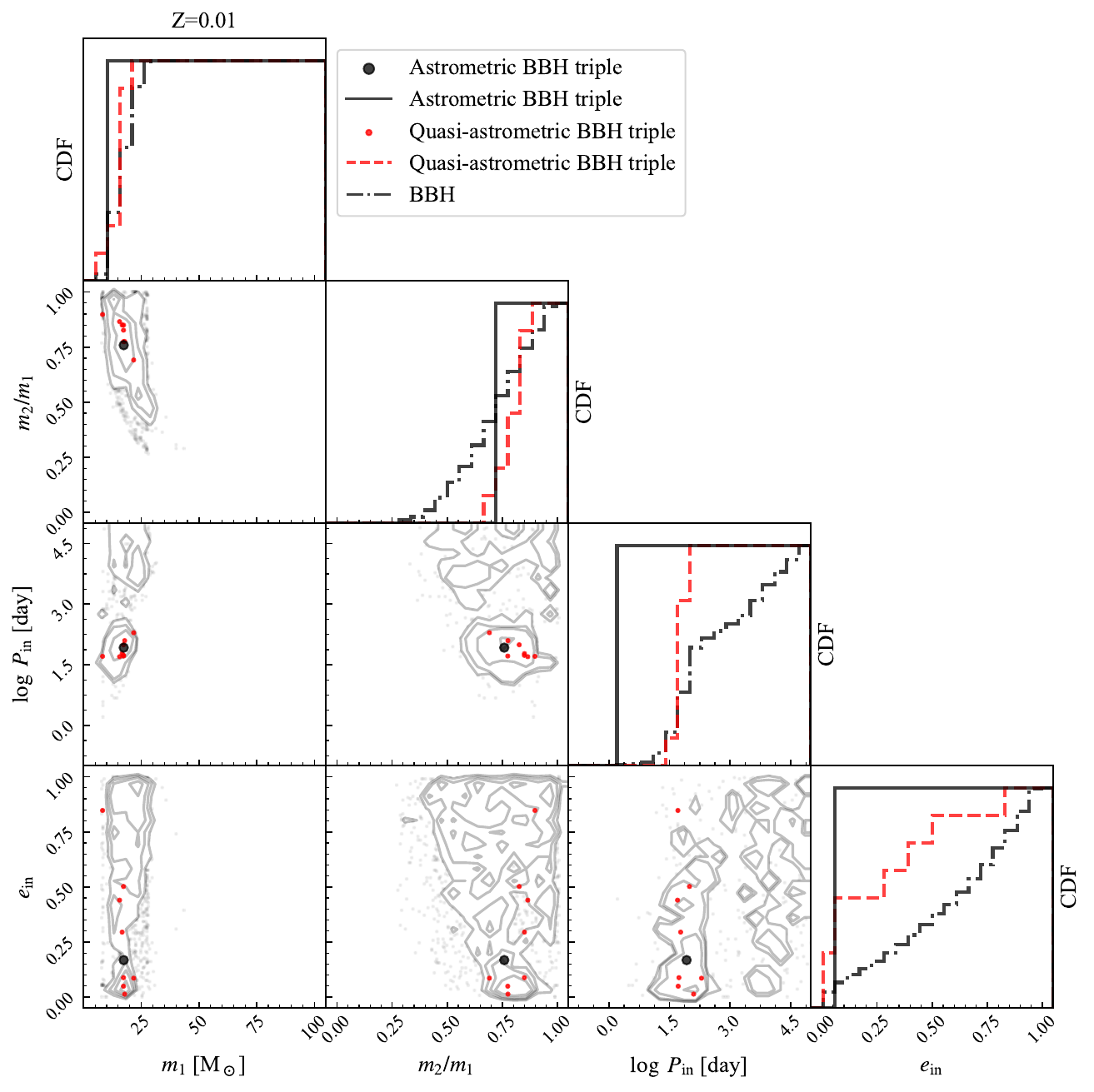}
  \caption{The same as Figure \ref{fig:cornerplotAllbbh_metl_z2e-4}
    but for the $Z=0.01$ model.}
  \label{fig:cornerplotAllbbh_metl_z1e-2}
\end{figure*}

\begin{figure*}
  \includegraphics[width=\columnwidth]{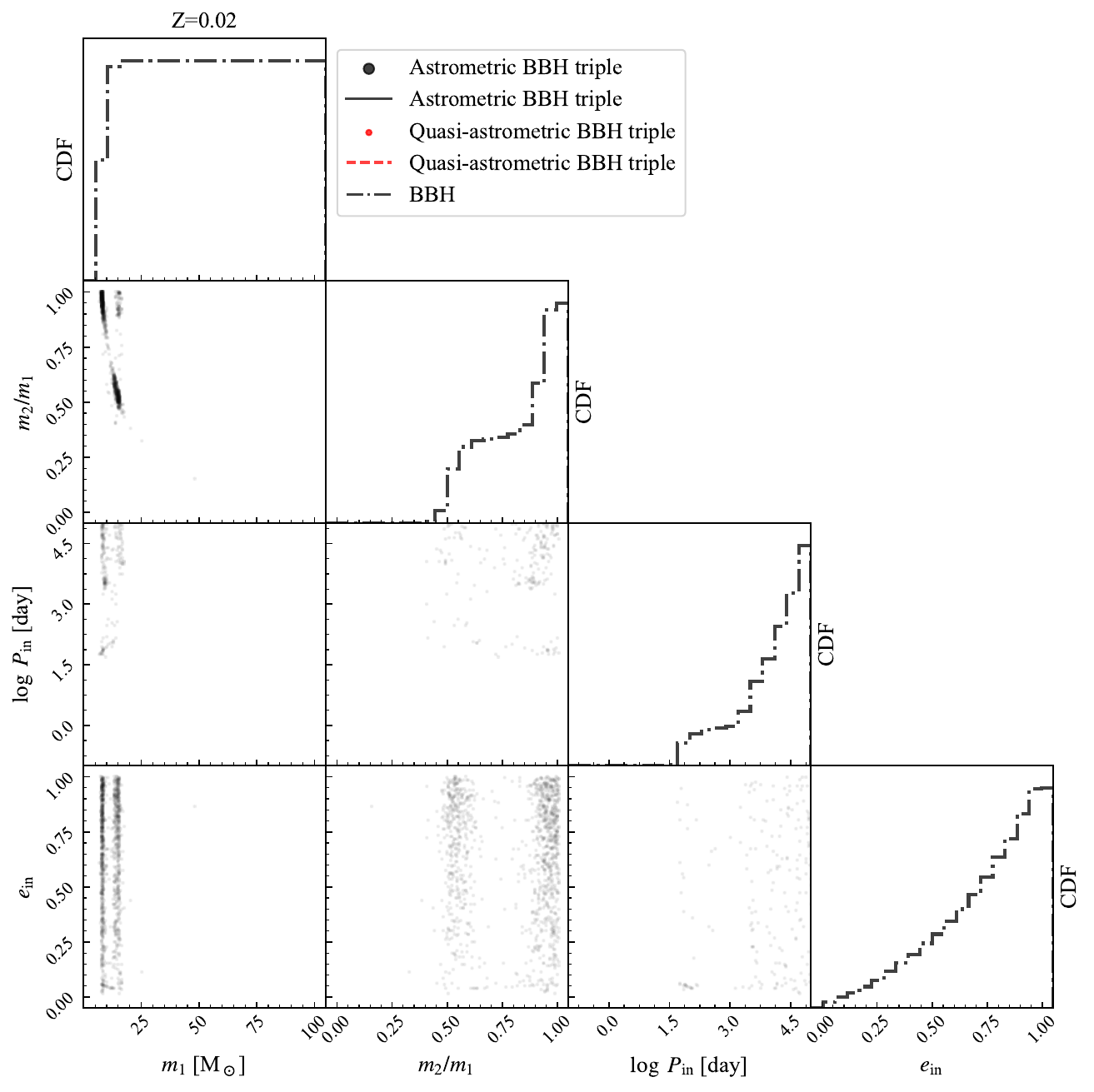}
  \caption{The same as Figure \ref{fig:cornerplotAllbbh_metl_z2e-4} but
    for the $Z=0.02$ model. No \gaia BBH triples or quasi-\gaia
    BBH triples are formed in this model, so their histograms and
    points are not plotted.  Furthermore, the number of all BBHs is
    small so their contours are not generated.}
  \label{fig:cornerplotAllbbh_metl_z2e-2}
\end{figure*}

Figures
\ref{fig:cornerplotGaiabbh_metl_z2e-3}--\ref{fig:cornerplotGaiabbh_metl_z2e-2}
and Figures
\ref{fig:cornerplotAllbbh_metl_z2e-3}--\ref{fig:cornerplotAllbbh_metl_z2e-2}
show the distributions of orbital parameters of \gaia BBH triples and
the inner BBHs of \gaia BBH triples, respectively, in all models other
than the $Z=0.0002$ model. The dark object masses of \gaia BH binaries
and \gaia BBH binaries increase with decreasing metallicity, as
typically observed in star and binary evolution models.

\section{Alternative binary evolution model}
\label{sec:alternative}

As shown in Figure \ref{fig:etaGaiabbh}, the formation efficiency of
\gaia BBH triples is zero at $Z=0.02$., the metallicity of typical
open clusters in the Milky Way galaxy
\citep{2010ARA&A..48..431P}. Thus, our simulation results indicate
that no \gaia BBH triples have formed in Milky Way open clusters at
the present time. Note that we need not consider open clusters outside
the Milky Way galaxy, because \gaia BH binaries will be found within
only 10 kpc (within the Milky Way galaxy) even by Gaia DR4 and Final
DR \citep{2025PASP..137d4202N}.

However, the formation efficiency of \gaia BBH triples at $Z=0.02$
strongly depends on the adopted binary evolution model, as described
in subsection \ref{sec:InnerBBH}. In an alternative binary evolution
model that admits common envelope evolution in the Hertzsprung gap
phase of a donor star, short-period BBHs can be formed at $Z=0.02$
\citep[e.g.][]{2020A&A...636A.104B} and the formation efficiency of
\gaia BBH triples at $Z=0.02$ may be comparable to those at $Z \le
0.005$; that is, $10^{-6}\;\msun^{-1}$. Assuming that the formation
efficiencies of \gaia BBH triples are $10^{-6}\;\msun^{-1}$ at all
metallicities, Milky Way open clusters can yield \gaia BBH triples at
the present time.  This alternative binary evolution model is
described below.

We first estimate the probability of discovering \gaia BBH triples
inside open clusters. We suppose that the formation rate of open
clusters (CFR) in the Milky Way galaxy has remained constant during
the past $10$ Gyrs. The $\tgw$ distribution of \gaia BBH triples is
modeled as
\begin{align}
  f(\tgw) = \left\{
  \begin{array}{ll}
    0 & (\tgw/{\rm yr} \le 10^7)\\
    1 & (10^7 \le \tgw/{\rm yr} \le 10^8)\\
    (\tgw/10^8\;{\rm yr})^{-1} & (10^8 \le \tgw/{\rm yr} \le
    10^{13})\\
    0 & (\tgw/{\rm yr} \ge 10^{13})
  \end{array}
  \right..
\end{align}
The $\tgw$ distribution over $\tgw \le 10^{10}$ yrs is based on the
results of \cite{2012ApJ...759...52D} and
\cite{2020MNRAS.498..495D}. As shown in Figure
\ref{fig:mergertimeGaiabbh}, the logarithmic flat distribution ends at
$\tgw = 10^{13}$ yrs. The number of \gaia BBH triples within $t_0 \le
\tgw \le t_1$ is determined as
\begin{align}
  N(t_0,t_1) = {\rm CFR} \; \int_{t_0}^{t_1} f(\tgw)
  \max(\tgw,10^{10}\;{\rm yr}) \; d\tgw.
\end{align}
Here, we assume that \gaia BBH triples are formed at the same time as
their host cluster and remain in the host until the host clusters are
disrupted. In other words, \gaia BBH triples are located in open
clusters while those clusters remain alive. Under this assumption, the
probability is overestimated. As the lifetime of open clusters is $1$
Gyrs, the fraction of \gaia BBH triples inside open clusters can be
written as
\begin{align}
  N(0,10^9)/N(0,10^{13}) \sim 0.012.
\end{align}
This fraction is smaller than the ratio of the cluster lifetime ($1$
Gyrs) to the Hubble time ($10$ Gyr), because most \gaia BBH triples
contain inner BBHs with $\tgw > 10^{10}$ Gyrs. Ultimately, the
probability of finding \gaia BBH triples inside open clusters is only
$\sim 1$ \% even in the alternative binary evolution model, where the
probability is overestimated.

We now assess the detectability of \gaia BBH triples by LISA and
TianQin. The number of \gaia BBH triples detectable by these
instruments is
\begin{align}
  N_{\rm det}(t_0,t_1) = {\rm CFR} \; \int_{t_0}^{t_1} f(\tgw)
  10^{7}\;{\rm yr} \; d\tgw,
\end{align}
where \gaia BBH triples are limited to those with $t_0 \le \tgw \le
t_1$. For an optimistic estimate, we place these \gaia BBH triples at
$1$ kpc where they are detectable in $10^7$ yrs. Note that when the
distance increases to $10$ kpc, the detectable duration reduces to
$10^6$ yrs and the detectability becomes smaller. The detection
probability can be written as
\begin{align}
  \frac{N_{\rm det}(0,10^{10})}{N(0,10^{13})} \sim 0.0011,
\end{align}
similar to that estimated in subsection
\ref{sec:gravitationalwave}. Assuming that \gaia BBH triples with
$\tgw \le 10^9$ are inside open clusters as described above, the
detection probability of \gaia BBH triples inside open clusters is
calculated as
\begin{align}
  \frac{N_{\rm det}(0,10^{9})}{N(0,10^{13})} \sim 0.0011,
\end{align}
which nearly equals the detection probability of all \gaia BBH
triples. Thus, if an \gaia BBH triple is detected by LISA and TianQin,
it will likely occupy an open cluster. However, we remark that the
detection probability is small, only $\sim0.1$\%.


\end{document}